\documentclass[twocolumn]{aastex701}

\usepackage{amsmath}
\usepackage{placeins}
\begin{document}

\title{Binding Energies of Astrophysically Relevant Molecules on CO$_2$ Clusters: Benchmarking and the Donor–Acceptor Origin of Surface-Dependent Binding}

\correspondingauthor{Aneesa Ahmad}
\author[0009-0005-6138-492X]{Aneesa Ahmad}
\affiliation{School of Physics and Astronomy, University of Leeds, Leeds, UK}
\email{a.ahmad@leeds.ac.uk}

\correspondingauthor{Catherine Walsh}
\author[0000-0001-6078-786X]{Catherine Walsh}
\affiliation{School of Physics and Astronomy, University of Leeds, Leeds, UK}
\email{c.walsh1@leeds.ac.uk}

\author[0000-0002-3102-1774]{Stefan Vogt-Geisse}
\affiliation{School of Physics and Astronomy, University of Leeds, Leeds, UK}
\affiliation{Departamento de Físico-Química, Facultad de Ciencias Químicas, Universidad de Concepción, Concepción, Chile}
\email{stvogtgeisse@qcmmlab.com}

\begin{abstract} 
Carbon dioxide ($\mathrm{CO_2}$) is among the most abundant interstellar ices and has been observed by JWST in water-rich mixtures and pure $\mathrm{CO_2}$ environments. Astrochemical models adopt binding energies (BEs) measured on water ice because few BEs are available for $\mathrm{CO_2}$ ice. We calculate BE distributions for twelve adsorbates ($\mathrm{N_2}$, $\mathrm{CH_4}$, $\mathrm{HCl}$, $\mathrm{C_2H_2}$, $\mathrm{H_2S}$, $\mathrm{HF}$, $\mathrm{HCN}$, $\mathrm{H_2CO}$, $\mathrm{HCOOH}$, $\mathrm{HNC}$, $\mathrm{CH_3OH}$, $\mathrm{H_2O}$) on $\mathrm{CO_2}$ clusters containing two to five molecules using DFT calibrated against CCSD(T)/CBS, a coupled-cluster treatment extrapolated to the complete-basis-set limit. Binding strength depends on whether the adsorbate requires the surface to donate a hydrogen bond (H-bond). $\mathrm{CO_2}$ can accept an H-bond but cannot donate one, and all six species compared with water ice from \citet{ahmad2026} bind more weakly on $\mathrm{CO_2}$. The ratio of the BE on $\mathrm{CO_2}$ to water ice ranges from 0.66 for $\mathrm{H_2CO}$ and 0.75 for $\mathrm{H_2O}$ to 0.87 for $\mathrm{CH_3OH}$, 0.89 for $\mathrm{HCOOH}$ and 0.92 for $\mathrm{C_2H_2}$. Formaldehyde is predicted to desorb near 57\,K from $\mathrm{CO_2}$-rich ice, compared with 85\,K from water-rich ice, and could contribute to cold gas-phase $\mathrm{H_2CO}$. Although the mean binding energies are changing with cluster size for several species, the differences can be traced to local binding interactions likely to remain important in larger ice models. The BEs vary by more than 500\,K between sites, while the desorption peaks range from 11\,K for $\mathrm{CH_4}$ to 132\,K for $\mathrm{HCOOH}$. The limited H-bonding capacity of $\mathrm{CO_2}$ may favour separation of mixed $\mathrm{H_2O}$:$\mathrm{CO_2}$ ice into $\mathrm{CO_2}$-rich domains.
\end{abstract}
\section{\textbf{Introduction}} \label{sec:intro}
In the cold, dense regions of molecular clouds and protoplanetary disks, volatile molecules freeze onto dust grain surfaces, building up chemically layered ice mantles \citep{Boogert_2015}. Among these, CO$_2$ is one of the most abundant ice constituents \citep{Pontoppidan_2008, Boogert_2015}. The James Webb Space Telescope (JWST) is now detecting CO$_2$ ice with unprecedented sensitivity and spatial resolution across all stages of star and planet formation, from molecular clouds to protoplanetary disks \citep{Mcclure_2023, Sturm_2023, Brunken_2024, bergner2024, potapov2025, bergner2026}. Mapping of lines of sight through the Chamaeleon I molecular cloud puts the CO$_2$/H$_2$O column density ratio at $\sim 0.15$ where the dust column is lowest, rising to $\sim 0.4$ towards the densest regions, with CO$_2$ making up 15--25\% of the total H$_2$O:CO$_2$:CO ice inventory \citep{Smith2025}. These observations reveal that CO$_2$ is found in both mixed and pure ices, reflecting the thermal history of the environment. Pure CO$_2$ ice forms through thermal processing. At temperatures of 20--30\,K, CO evaporates from the CO:CO$_2$ ice matrix through distillation, leaving CO$_2$ behind \citep{Pontoppidan_2006, Pontoppidan_2008}, while at higher temperatures CO$_2$ diffuses through and segregates from the H$_2$O-rich ice matrix into pure CO$_2$ regions. This segregation has been measured in the laboratory \citep{Oberg_2009, He_2017, Cooke_2018}, where it becomes efficient between $\sim$50 and 80\,K but is expected to occur at lower temperatures in space, over longer timescales \citep{Escribano_2013}.

These CO$_2$-rich domains represent a distinct surface environment whose desorption properties cannot be inferred from those of H$_2$O ice. H$_2$O ice can donate and accept hydrogen bonds (H-bonds), whereas CO$_2$ can only accept: a molecule bound to water ice may be held partly by a H-bond donated back to it by the surface, a contribution that is absent on CO$_2$. How strongly a molecule is held, its binding energy (BE), therefore depends on the surface. It both determines the temperature at which a species desorbs and the radius in the disk at which it transitions from ice to gas: its snowline \citep{Hasegawa_1992, Collings2004}. Snowline locations shape the chemical composition of forming planets by controlling where volatile species are available in the gas phase versus locked in ice \citep{Oberg_2011, Piso_2015, _berg_2021}. 

CO$_2$ ice is identified through its vibrational absorption bands at near- and mid-infrared wavelengths, most notably the 15.2$ \mu$m bending mode \citep{Pontoppidan_2008, Boogert_2015}. The profile of this feature is sensitive to the local molecular environment: when CO$_2$ is embedded in a H$_2$O-rich matrix, the band appears broader and redshifted, reflecting interactions with surrounding water molecules whereas in heated environments near protostars, the feature sharpens and becomes more symmetric, a signature of pure CO$_2$ ice \citep{Pontoppidan_2008, Escribano_2013}. Therefore CO$_2$ resides in multiple chemical environments simultaneously. Recent JWST mapping of quiescent molecular clouds finds the CO$_2$ to be predominantly mixed with water ice \citep{Mcclure_2023, Smith2025}, consistent with CO$_2$ forming alongside water early in a cloud's evolution, before any thermal processing. Once formed, this mixed CO$_2$ remains trapped in the ice well above the desorption temperature of pure CO$_2$ \citep{Fayolle_2011}.

Beyond spectroscopic observations, a few laboratory studies have directly investigated molecular desorption from CO$_2$ ice. \citet{Luna_2008} measured the adsorption of CH$_4$ onto solid CO$_2$ using quartz-crystal microbalance measurements, identifying desorption peaks consistent with physisorption, binding through weak van der Waals and electrostatic forces rather than chemical bonds, though without deriving quantitative BE values or accounting for site-to-site variability. Several temperature programmed desorption (TPD) experiments have since examined how CO$_2$ ice traps CO. \citet{Cooke_2018} measured the barrier for CO to diffuse into CO$_2$ ice, \citet{Simon_2019} showed that CO$_2$ ice holds CO more effectively than amorphous solid water (ASW) up to the onset of CO$_2$ desorption near 70\,K, with implications for the gas/solid partitioning of volatile species in disk regions where CO$_2$ dominates the grain surface, and \citet{Pesciotta_2024} found this trapping to be efficient across a range of mixing ratios and ice thicknesses. Together, these studies confirm that CO$_2$ surfaces play a distinct role in volatile retention, but their coverage remains limited to individual species under specific conditions. At the molecular level, complexes of individual species with one or two CO$_2$ molecules have been characterised by matrix-isolation infrared spectroscopy \citep{Soulard_2015, Soulard_2023} and by ab-initio calculation \citep{Danten_2005, Thanthiriwatte_2012}, largely for atmospheric or industrial applications rather than for interstellar ices. To our knowledge, BEs have not been reported for a range of astrophysically relevant species on CO$_2$ clusters of increasing size, and astrochemical models continue to rely on ASW-derived values regardless of the actual grain-surface composition \citep{Cuppen_2017, Minissale_2022}.

In this work we present BE distributions for twelve astrophysically relevant adsorbates on CO$_2$ clusters, computed with the Binding Energy Evaluation Platform \citep[BEEP;][]{Bovolenta_2022}\footnote{\url{https://github.com/QCMM/beep}}. First, the cluster geometries from our production method were compared against a higher level MP2 wavefunction reference. MP2 (second-order M{\o}ller--Plesset perturbation theory) is a quantum chemical method more accurate than standard density functional theory (DFT) but more computationally expensive, making it a useful reference for validating cheaper methods on small systems. Second, the BEs from 114 density functionals are tested against a CCSD(T) reference (coupled-cluster theory), the most accurate practical method for molecules of this size, extrapolated to the complete basis set limit (CBS), an estimate of the exact result with this source of numerical error removed. We adopt $\omega$B97X-V for all reported values. Tracking the mean BE with cluster size identifies which species show a binding energy that has begun to level off at $n = 5$ and which are still rising. Finally we convert the BEs into desorption temperatures and compare them with our previous results on ASW \citep{ahmad2026}.
\section{\textbf{Computational Methodology}}
\label{sec:comp_method}

\subsection{Cluster Models}
\label{sec:cluster_models}

\begin{figure}
    \centering
    \includegraphics[width=3.1in, height=4.3in]{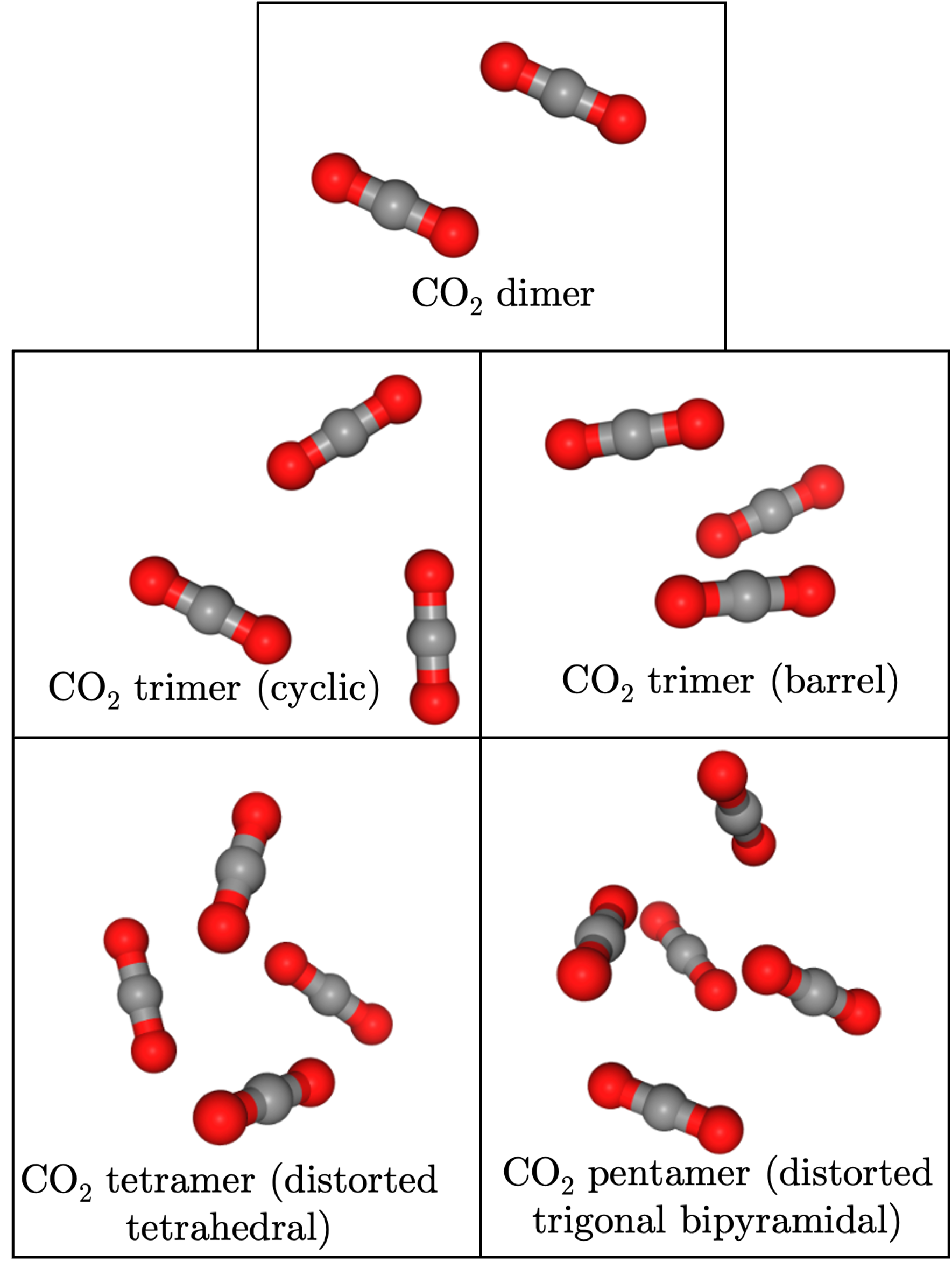}
    \caption{\centering Carbon dioxide (CD; CO$_2$) clusters modelled in this study (Dimer, CD${_2}$-01; cyclic trimer, CD${_3}$-01; barrel trimer, CD${_3}$-02; tetramer, CD${_4}$-01; pentamer, CD${_5}$-01).}
    \label{fig:CO2}
\end{figure}

To probe the size dependence of BEs with CO$_2$ cluster size, we model the adsorbate--CO$_2$ interaction on a series of small CO$_2$ clusters of size $n = 2$--$5$, where $n$ is the number of CO$_2$ molecules. The geometries used here are the established low-energy minima of the $(\mathrm{CO_2})_n$ potential energy surface, determined by high-resolution spectroscopy and ab initio calculation \citep{Jucks_1987, jovan2008, takeuchi2008}. Larger pure CO$_2$ clusters have been characterised by electron diffraction \citep{Torchet_1984}, by molecular dynamics \citep{Cardini_1989}, and at the MP2 level up to $n = 13$ \citep{Yeole_2012}. Of the low-energy minima in literature, five cluster geometries are employed for the binding-site analysis in this study (Figure~\ref{fig:CO2}): the dimer (CD${_2}$-01), two trimer isomers, the cyclic (CD${_3}$-01) and barrel (CD${_3}$-02) configurations, the distorted-tetrahedral tetramer (CD${_4}$-01), and the distorted-trigonal-bipyramidal pentamer (CD${_5}$-01). The isolated CO$_2$ monomer (CD${_1}$-01) does not constitute a meaningful ice analogue and is therefore excluded from the binding-site analysis; it is used only as the substrate for the geometry consistency check (Appendix~\ref{sec:geometry_consistency_method}).
Initial cluster geometries were constructed manually in Avogadro \citep{hanwell2012}\footnote{\url{https://avogadro.cc/}} and then geometry-optimised within the BEEP workflow described in Section~\ref{sec:geometry_refinement}.

Twelve astrophysically relevant adsorbates were studied on each cluster: N$_2$, CH$_4$, H$_2$CO, HCl, C$_2$H$_2$, H$_2$S, HF, HCN, HCOOH, HNC, CH$_3$OH, and H$_2$O. These include various binding behaviour on a $\mathrm{CO_2}$ surface, from weak dispersion-bound species to strong H-bond donors. The species were divided into three broad groups: non-donors bound mainly by dispersion ($\mathrm{N_2}$ and $\mathrm{CH_4}$); weak donors with modest dipoles or mildly positive hydrogen atoms (HCl, $\mathrm{C_2H_2}$, $\mathrm{H_2CO}$, and $\mathrm{H_2S}$); and strong H-bond donors (HF, HCN, HNC, HCOOH, $\mathrm{CH_3OH}$, and $\mathrm{H_2O}$).
\subsection{Geometry Sampling and Refinement}
\label{sec:geometry_refinement}

\subsubsection{BEEP Sampling and Geometry Refinement}
\label{subsec:beep_workflow}

Binding sites were generated and refined using the Binding Energy Evaluation Platform \citep[BEEP;][]{Bovolenta_2022}, version 0.6, with the updated protocol described in \citet{ahmad2026}. The BEEP workflow proceeds in three stages: (i) binding-site sampling, in which adsorbate molecules are placed on the cluster surface in randomised orientations and positions and subjected to a coarse geometry relaxation; (ii) geometry refinement of the geometrically unique sampled sites at a production level of theory; and (iii) computation of the harmonic Hessian for the zero-point vibrational energy (ZPVE) correction and assembly of the ZPVE-corrected BE distribution. Initial geometry optimisations were performed at BLYP-D3BJ/def2-SVP \citep{becke1988,lee1988,Grimme2010}. Duplicate configurations were removed using a root-mean-square deviation (RMSD) criterion of 0.4\,\AA{}, so that near-identical structures are counted as a single binding site. The resulting binding site geometry was improved for each unique species using MPWB1K-D3BJ \citep{Zhao2004} with a def2-SVPD \citep{Rappoport2010} basis set. The performance of the MPWB1K-D3BJ/def2-SVPD refinement against the MP2 reference is established in Appendix~\ref{sec:geometry_consistency}.
\subsubsection{Binding-site seeding for larger clusters}
\label{subsec:binding_site_seeding}

For the larger clusters (CD4 and CD5), the BEEP sampling was complemented with additional binding-site configurations derived from the smaller clusters (CD2 and CD3). Binding arrangements identified on the smaller clusters were transferred to the larger clusters and subsequently re-optimised. This procedure provided an expanded set of chemically meaningful binding configurations for the larger clusters and facilitated a more consistent comparison of binding motifs across the cluster series.

For a target cluster CD$n$, the relative adsorbate--CO$_2$ geometry of each converged binding site on a smaller cluster CD$m$ ($m < n$) was superimposed onto a CO$_2$ unit of the CD$n$ cluster, generating a new trial structure that was then subjected to the full BEEP refinement protocol described in Section~\ref{subsec:beep_workflow}. Seed structures are labelled in the sample lists by their source: SEED\_CD3 denotes a site seeded from a converged CD3 binding geometry, while CD$n$\_..\_SEED\_V2 denotes a second seeding round drawing on CD$m$ ($m = 2, 3$) geometries. This procedure preserves chemically sensible adsorbate geometries (e.g.\ the H-bond donor/acceptor alignment established on the smaller cluster) while allowing the additional CO$_2$ units of the larger cluster to reorganise around the adsorbate during optimisation.

The seeding strategy works for two reasons. Each molecule binds through a single local contact, so an arrangement found on a small cluster still holds on a larger one, letting seeding start from a sensible geometry, consistent with the small non-additive contribution found for H$_2$O with two CO$_2$ molecules \citep{Danten_2005}. And because seeded and sampled structures are pooled with duplicates merged, any seeded structure matching an existing one is discarded, so seeding only contributes additional sites.

Seeded and natively sampled sites are pooled and duplicates were removed as described in Section~\ref{subsec:beep_workflow}, using a heavy-atom Kabsch RMSD threshold of 0.5\,\AA{}, compared with the 0.4\,\AA{} all-atom criterion applied to the raw sampled geometries. The set of distinct minima is insensitive to this choice: the pairwise RMSD distribution among optimised structures is strongly bimodal, with duplicate pairs falling below 0.34\,\AA{} and the next-closest distinct pair at 0.65\,\AA{}. Any threshold between approximately 0.4 and 0.6\,\AA{} therefore recovers the same set of sites (16 distinct minima for HF on CD4), and an all-atom rather than heavy-atom RMSD gives identical results.
\subsubsection{Treatment of Unconverged Sites}
\label{subsec:rescue_protocol}

For the following six adsorbates on the pentamer cluster, N$_2$, CH$_4$, H$_2$CO, HCl, H$_2$O, and C$_2$H$_2$, geometry optimisation at the MPWB1K-D3BJ/def2-SVPD refinement level reached our convergence criteria at only one of the 177 sampled and seeded sites. This is a known difficulty with functionals of this type on very flat potential energy surfaces (PES), where a weakly bound molecule can sit in many nearly equivalent orientations of almost the same energy and the optimisation oscillates between them without settling. The single surviving structure, a C$_2$H$_2$ site, leaves too few minima to characterise any of these systems, so we re-optimised all six sets of sites with a different, more numerically stable functional (BLYP-D3BJ/def2-SVPD) that converges reliably on such surfaces.

For these sites, geometries were optimised in Cartesian coordinates to avoid a separate numerical problem that can affect near-linear molecules. The def2-SVPD basis was retained throughout, to isolate the change in functional. BEs were then computed at the same three functionals used throughout ($\omega$PBE-D3BJ, $\omega$B97X-V, M06-HF; def2-TZVP) on these new geometries. The counterpoise (CP) correction was applied to remove the artificial over-binding caused by the two molecules sharing basis functions, known as the basis set superposition error (BSSE) \citep{Boys_1970}. The ZPVE corrections were taken from BLYP-D3BJ/def2-SVPD frequencies at the same geometry. Section~\ref{subsec:rescue_validation} quantifies the effect of this change of functional, which we carry as an additional uncertainty of $\pm122$\,K on these sites.

Of the 177 sampled and seeded sites, five lacked a converged starting geometry and could not be reoptimised, leaving 172. Two subsequent optimisations and one Hessian calculation failed to converge. We also excluded 34 structures in which the adsorbate was no longer bound to the cluster and eight in which a proton had transferred to a $\mathrm{CO_2}$ oxygen, since neither represents a physisorption site. The closest adsorbate--cluster contacts clearly distinguished these structures from the retained sites, with no ambiguous cases. A further 11 sites were removed because they carried imaginary frequencies, leaving 116 sites.
\subsubsection{Validation of BLYP-D3BJ/def2-SVPD}
\label{subsec:rescue_validation}

To verify the use of BLYP-D3BJ/def2-SVPD in place of MPWB1K, we reoptimised 24 sites for which both methods converged. The test set covers eight adsorbates on the dimer, both trimers and the tetramer, the clusters on which MPWB1K converged reliably. For each relaxed structure, we recalculated the CP-corrected BE and $\Delta$ZPVE (Appendix Table~\ref{tab:rescue_protocol_validation}). This comparison tests the full protocol because changing the functional affects both the optimised geometry and the corresponding vibrational correction. In each case, the Hessian was calculated using the same method as the geometry optimisation.

The two methods produce similar geometries. For the tetramer, the mean RMSD between the MPWB1K and BLYP minima is $0.105$\,\AA{} for the 18 of 20 control sites at which both methods located the same basin (the two exceptions are among the three basin changes discussed below). The median difference in the CP-corrected BE is $10$\,K, while the mean difference in $\Delta$ZPVE is $86$\,K. After both contributions are included, the final $E_\mathrm{b}$ values differ by $122$\,K on average, with a median difference of $58$\,K. We use the mean difference to assign an additional uncertainty of $\pm122$\,K to the rescued sites. This uncertainty applies only to pentamer sites for the six adsorbates treated using the BLYP level. 

In three of the 24 control calculations, BLYP converged to a different local binding geometry, giving an RMSD greater than $0.3$\,\AA{}. If these cases are excluded, the mean difference in the final $E_\mathrm{b}$ falls to $69$\,K. We nevertheless include them when estimating the additional uncertainty because a similar change in binding geometry cannot be ruled out for the rescued sites, where no converged MPWB1K result is available for comparison.

For the two weakest-binding species, $\mathrm{N_2}$ and $\mathrm{CH_4}$, the additional uncertainty corresponds to $22\%$ and $37\%$ of the tabulated $E_\mathrm{b}$, respectively. In both cases, it remains smaller than the variation between binding sites, which is also the larger source of variation for the other four species calculated with BLYP. None of the BLYP sites involve the strong H-bond donors $\mathrm{HCOOH}$, $\mathrm{HF}$, $\mathrm{CH_3OH}$, $\mathrm{HCN}$, or $\mathrm{HNC}$, so the size trends reported for them are unaffected. H$_2$O is the exception among the strong donors: its pentamer sites were computed on BLYP-D3BJ/def2-SVPD geometries, so its tetramer-to-pentamer step spans the change of method.
\subsection{Binding Energy Calculation}
\label{sec:be_calculation}

CP-corrected BEs \citep{Boys_1970} were computed at three density functionals with the def2-TZVP basis: $\omega$PBE-D3BJ \citep{Perdew1996, vydrov2006, Grimme_D3BJ}, $\omega$B97X-V \citep{mardirossian2014}, and M06-HF \citep{zhao2006}. The final reporting level BE is the $\omega$B97X-V/def2-TZVP, which is calibrated against a reference-quality CCSD(T)/CBS dataset in Appendix~\ref{sec:be_benchmark}; the remaining two functionals are retained as a methodological cross-check. The BE of an adsorbate--cluster complex is defined as
\begin{equation}
  \Delta E_\mathrm{b} = -\left(E_{X+\mathrm{ice}} - \left(E_{X} + E_\mathrm{ice}\right) + \Delta\mathrm{ZPVE}\right),
\end{equation}
with the ZPVE correction
\begin{equation}
  \Delta\mathrm{ZPVE} = E_\mathrm{ZPVE,\,X+\mathrm{ice}} - \left(E_\mathrm{ZPVE,\,X} + E_\mathrm{ZPVE,\,ice}\right),
\end{equation}
where $E_{X+\mathrm{ice}}$ is the energy of the adsorbate--cluster complex, $E_{X}$ is the energy of the isolated adsorbate, and $E_\mathrm{ice}$ is the energy of the isolated CO$_2$ cluster. The negative sign follows the convention that $\Delta E_\mathrm{b} > 0$ for a bound adsorbate. Per-site $\Delta$ZPVE values were obtained from direct complex Hessians for all sites in the dataset, computed at MPWB1K-D3BJ/def2-SVPD throughout, except for the pentamer sites of the six adsorbates listed in Section~\ref{subsec:rescue_protocol} (N$_2$, CH$_4$, H$_2$CO, HCl, H$_2$O and C$_2$H$_2$), for which the rescue level BLYP-D3BJ/def2-SVPD was used. In all cases the fragment ZPVEs were evaluated at the same level as the corresponding complex. Each Hessian was evaluated at a stationary point of the level of theory used to optimise its geometry. No regression prediction or mean fallback was applied to $\Delta$ZPVE for any species in the final dataset. 

All DFT calculations were performed with the Psi4 software package \citep{smith2020}\footnote{\url{https://psicode.org/}}.
\section{\textbf{Benchmarking}}
\label{sec:benchmark_main_body}

The BEs reported below were validated in two ways (Appendix~\ref{sec:benchmarking}). First, the production optimisation method, MPWB1K-D3BJ/def2-SVPD, was compared with an MP2/def2-TZVP reference for the twelve adsorbate--CO$_2$ monomer complexes. The geometries agree to within $0.1$\,\AA{} heavy-atom RMSD for nine of the ten adsorbates that converged, with a median RMSD of $0.034$\,\AA{}. The only outlier, HCOOH, reflects two nearly equivalent monomer orientations and does not persist for the larger clusters used to determine the BEs (Appendix~\ref{sec:geometry_consistency}). Second, 114 density functionals were compared with a CCSD(T)/CBS reference across 36 systems covering all twelve species. $\omega$B97X-V/def2-TZVP gives the best agreement (ME $= +6$\,K, MAE $= 53$\,K, RMSE $= 72$\,K) and is therefore used for all reported BEs. Its MAE sets the minimum uncertainty $\sigma$ quoted throughout Section~\ref{sec:BEs} (Appendix~\ref{sec:be_benchmark}). Recomputing with the two additional functionals reproduces the same tetramer-to-pentamer trend for ten of the twelve adsorbates. The exceptions are $\mathrm{N_2}$, for which $\omega$PBE-D3BJ and M06-HF give small decreases ($-13$ and $-30$\,K) where $\omega$B97X-V gives a rise of 77\,K, and HCl, for which M06-HF gives a decrease of 97\,K where the other two give rises of 57 and 55\,K. All of these changes are below 100\,K and smaller than the pentamer site-to-site spread, so the functionals agree that the step is small for both species and differ only in its sign.
\section{\textbf{Results \& Discussion}}
\label{sec:res_disc}
\subsection{Binding Energies}
\label{sec:BEs}

We report the full spread of BEs across sites, for each species and cluster size, rather than a single mean, because both ends of the distribution matter astrophysically: the strongest binding sites set the temperature at which a species stops thermally desorbing, while the weakest sites provide the easiest routes for a molecule to leave, either by desorbing or diffusing to a more strongly bound site. We quantify this spread by $\sigma_\mathrm{site}$, the standard deviation of the BE across the different sites on a single cluster, a measure of how much the binding strength varies with where the molecule sits on the cluster. These values are large (often $>500$\,K, and $>1000$\,K for the strongest H-bond donors), corresponding to tens of kelvin in desorption temperature, so reducing each species to a single BE would hide both the range of binding strengths the surface offers and its consequences for the composition of the ice mantle (Section~\ref{sec:astro_implications}).

\begin{figure*}
    \centering
    \includegraphics[width=7in, height=3.6in]{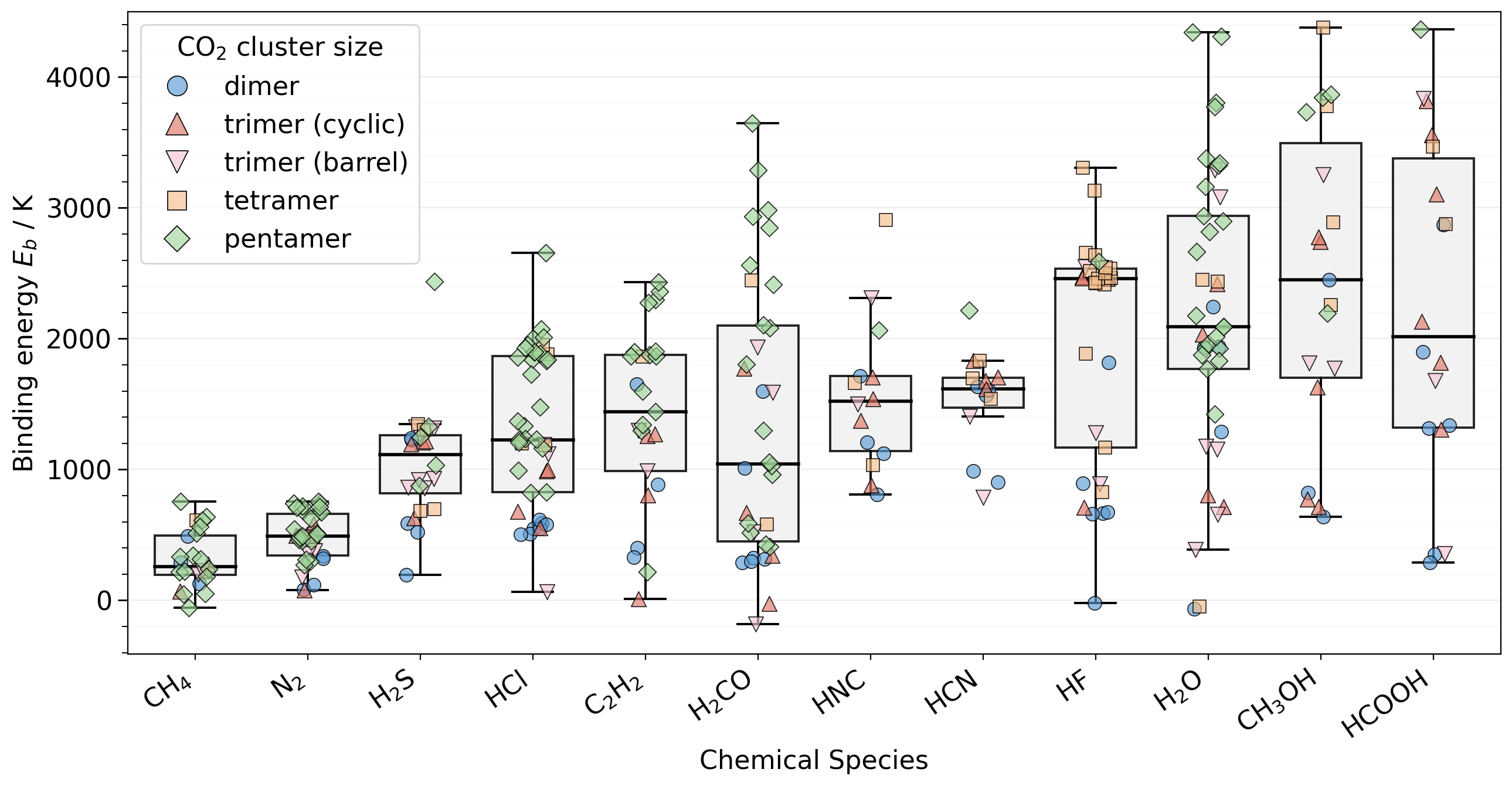}
    \caption{Per-species binding energy distributions on CO$_2$ clusters
    ($n = 2$--$5$) at the $\omega$B97X-V/def2-TZVP level of theory, where the horizontal black line is the median.}
    \label{fig:be_boxplots}
\end{figure*}

The BEs for the twelve adsorbates on CO$_2$ clusters of size $n = 2$--$5$ are given in Table~\ref{tab:final_BE}, with the per-cluster distributions shown in Figure~\ref{fig:be_boxplots}. All values are at $\omega$B97X-V/def2-TZVP (Appendix~\ref{sec:be_benchmark}). We quote $\sigma_\mathrm{site}$ as the uncertainty on each mean, with a floor of 53\,K, the benchmark error against the CCSD(T)/CBS reference. Trimer entries pool the sites from both the cyclic (CD${_3}$-01) and barrel (CD${_3}$-02) cluster shapes.
\begin{table*}[t]
\centering
\footnotesize
\renewcommand{\arraystretch}{1.15}

\caption{ZPVE-corrected mean binding energies ($E_{\mathrm{b}}$) in K of 12 astrophysically relevant adsorbates on $\mathrm{CO_2}$ clusters of size $n=2$--$5$.}
\label{tab:final_BE}

\begin{tabular*}{\textwidth}{@{\extracolsep{\fill}}lcccccc@{}}
\hline\hline
Species &
Dimer &
Trimer &
Tetramer &
Pentamer &
$\Delta E_{\mathrm{b}}(4\rightarrow5)$ &
Trend at $n=5$ \\
&
$E_{\mathrm{b}}\pm\sigma\,(n)$ &
$E_{\mathrm{b}}\pm\sigma\,(n)$ &
$E_{\mathrm{b}}\pm\sigma\,(n)$ &
$E_{\mathrm{b}}\pm\sigma\,(n)$ &
&
\\
\hline
$\mathrm{N_2}$     & $275\pm178\ (5)$ & $365\pm182\ (7)$ & $486\pm53\ (1)$ & $563\pm156\ (21)$ & $+77\ (+14\%)$ & rising\tablenotemark{*} \\
$\mathrm{CH_4}$    & $302\pm181\ (3)$ & $198\pm77\ (5)$ & $611\pm53\ (1)$ & $331\pm239\ (15)$ & $-280\ (-85\%)$ & non-monotonic\tablenotemark{*} \\
$\mathrm{HCl}$     & $557\pm53\ (7)$ & $799\pm397\ (7)$ & $1554\pm418\ (4)$ & $1612\pm451\ (25)$ & $+57\ (+4\%)$ & levelling \\
$\mathrm{C_2H_2}$  & $815\pm609\ (4)$ & $939\pm494\ (6)$ & $1865\pm53\ (1)$ & $1763\pm578\ (14)$ & $-102\ (-6\%)$ & non-monotonic\tablenotemark{*} \\
$\mathrm{H_2S}$    & $754\pm462\ (5)$ & $1046\pm235\ (10)$ & $1008\pm369\ (4)$ & $1385\pm615\ (5)$ & $+377\ (+27\%)$ & non-monotonic \\
$\mathrm{HF}$      & $783\pm596\ (6)$ & $2030\pm752\ (10)$ & $2368\pm622\ (16)$ & $2590\pm53\ (1)$ & $+222\ (+9\%)$ & levelling\tablenotemark{*} \\
$\mathrm{HCN}$     & $1340\pm361\ (5)$ & $1505\pm379\ (6)$ & $1690\pm147\ (3)$ & $2216\pm53\ (1)$ & $+526\ (+24\%)$ & rising steeply\tablenotemark{*} \\
$\mathrm{H_2CO}$   & $638\pm548\ (6)$ & $827\pm828\ (8)$ & $1512\pm1321\ (2)$ & $1831\pm1073\ (18)$ & $+319\ (+17\%)$ & rising \\
$\mathrm{HCOOH}$   & $1345\pm974\ (6)$ & $2401\pm1236\ (9)$ & $3172\pm418\ (2)$ & $4364\pm53\ (1)$ & $+1193\ (+27\%)$ & rising steeply\tablenotemark{*} \\
$\mathrm{HNC}$     & $1213\pm377\ (4)$ & $1553\pm466\ (6)$ & $1868\pm955\ (3)$ & $2065\pm53\ (1)$ & $+197\ (+10\%)$ & levelling\tablenotemark{*} \\
$\mathrm{CH_3OH}$  & $1303\pm997\ (3)$ & $1934\pm932\ (8)$ & $3327\pm939\ (4)$ & $3410\pm814\ (4)$ & $+84\ (+2\%)$ & levelling \\
$\mathrm{H_2O}$    & $1468\pm925\ (5)$ & $1572\pm1057\ (10)$ & $1613\pm1437\ (3)$ & $2689\pm864\ (23)$ & $+1076\ (+40\%)$ & rising steeply \\

\hline
\end{tabular*}

\vspace{0.5em}

\begin{minipage}{\textwidth}
\footnotesize
\textit{Note---}
The reported values are ZPVE-corrected BEs calculated at $\omega$B97X-V/def2-TZVP, with $\Delta$ZPVE from a direct Hessian at every site. Each entry gives the mean BE, its uncertainty, and the number of binding sites as $E_{\mathrm{b}}\pm\sigma\,(n)$. The uncertainty is the larger of the site-to-site standard deviation and the benchmark MAE of $53$\,K; a value of $53$\,K therefore denotes either a single retained site or a spread smaller than the benchmark error. The trimer results combine sites from the cyclic CD$_3$-01 and barrel CD$_3$-02 isomers. The pentamer values for N$_2$, CH$_4$, HCl, C$_2$H$_2$, H$_2$CO, and H$_2$O were computed on BLYP-D3BJ/def2-SVPD geometries (Section~\ref{subsec:rescue_protocol}) and carry an additional methodological uncertainty of $\pm122$\,K that is not included in $\sigma$. Six sites have $E_{\mathrm{b}}\le0$ because the zero-point correction exceeds a shallow electronic binding energy (CH$_4$ pentamer, $-58$\,K; H$_2$CO trimer, $-27$ and $-182$\,K; H$_2$O dimer, $-64$\,K; H$_2$O tetramer, $-46$\,K; HF dimer, $-20$\,K). All six are retained in all reported statistics. For such shallow minima the harmonic $\Delta$ZPVE, which is dominated by soft, anharmonic intermolecular modes, cannot reliably fix the sign of $E_{\mathrm{b}}$, and excluding sites on the basis of their corrected value would bias the affected means upwards. $\Delta E_{\mathrm{b}}(4\rightarrow5)$ gives the change from the tetramer to the pentamer, with the percentage calculated relative to the pentamer value. For the trend labels, changes below $10\%$ are described as levelling, those between $10\%$ and $20\%$ as rising, and those above $20\%$ as rising steeply, based on the unrounded percentages (HNC, at $+9.5\%$, is therefore levelling). Species that do not increase consistently with cluster size are labelled non-monotonic.

\vspace{0.5em}

$^{*}$The trend assessment rests on a single site at one or both cluster sizes and should be treated as indicative only.
\end{minipage}

\end{table*}

\subsubsection{Strong hydrogen-bond donors}
\label{sec:H_bond_donors}

\begin{figure}
    \centering
    \includegraphics[width=3.1in, height=3.7in]{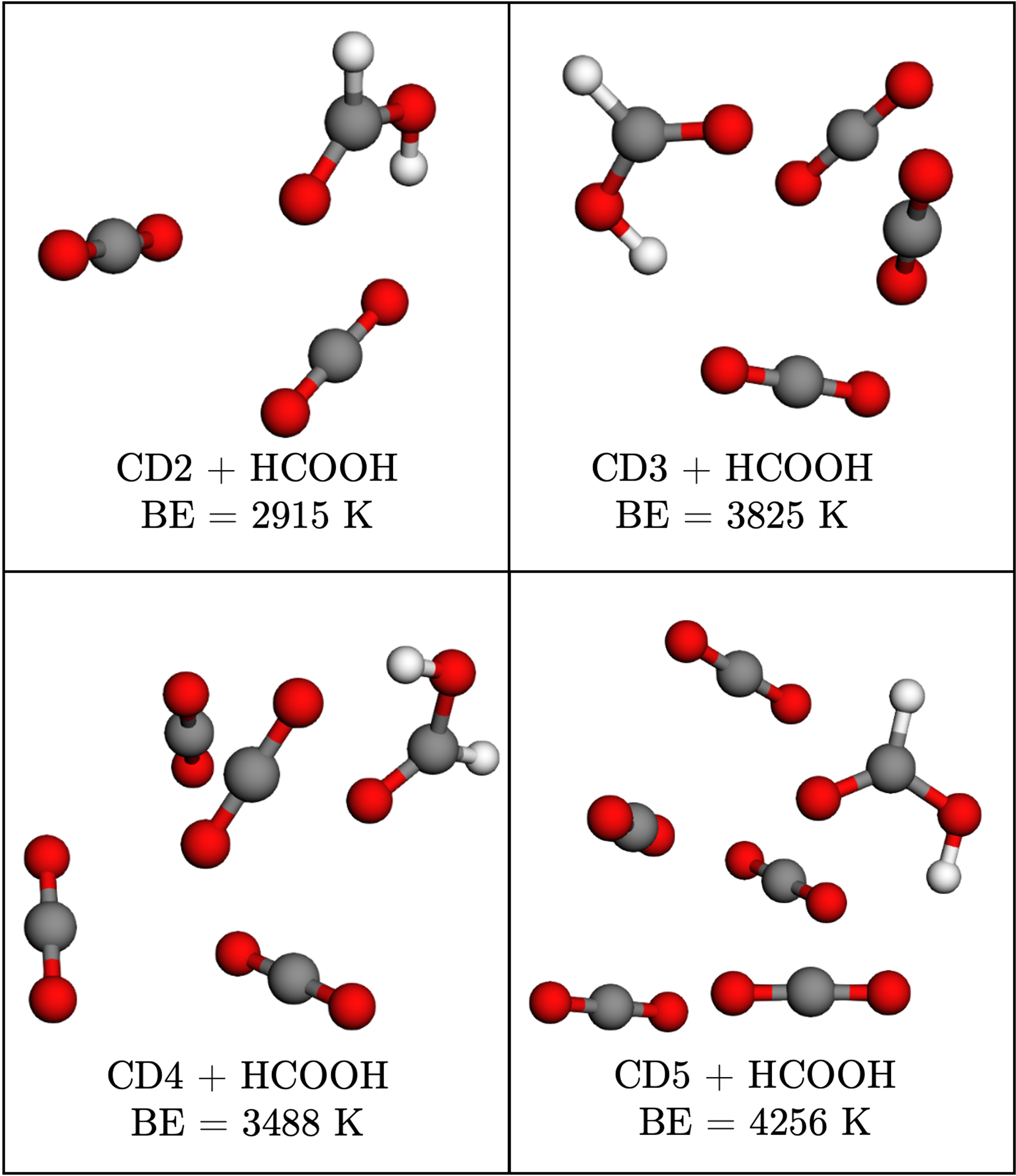}
    \caption{Representative HCOOH binding sites on CO$_2$ clusters of size
    $n = 2$--$5$ at $\omega$B97X-V/def2-TZVP, showing progressive
    engagement of additional CO$_2$ acceptors by the --COOH motif as
    cluster size grows.}
    \label{fig:hcooh}
\end{figure}
Six adsorbates (HCOOH, HF, CH$_3$OH, HCN, HNC, and H$_2$O) carry a polar X--H bond whose proton can donate a H-bond to one of the CO$_2$ oxygen atoms. For HCOOH, HF, and CH$_3$OH the average BE more than doubles from the dimer to the pentamer (Table~\ref{tab:final_BE}): HCOOH climbs from 1345\,K to 4364\,K (+225\%), HF from 783\,K to 2590\,K (+231\%), and CH$_3$OH from 1303\,K to 3410\,K (+162\%). HCOOH binds most strongly of all twelve adsorbates at the trimer and pentamer, because it can anchor to two CO$_2$ molecules at once: the O--H group donates a H-bond to one CO$_2$, while the carbonyl oxygen donates electron density to the electron-poor carbon of a neighbouring CO$_2$ (Figure~\ref{fig:hcooh}). On the isolated CO$_2$ monomer the production functional and the MP2 reference settle on different shallow HCOOH orientations (Appendix~\ref{sec:geometry_consistency_method}), but this only happens when HCOOH makes a single contact; from the dimer upward the two-CO$_2$ bridging geometry is well defined and the same across methods.
\begin{figure*}
    \centering
    \includegraphics[width=5.8in, height=7in]{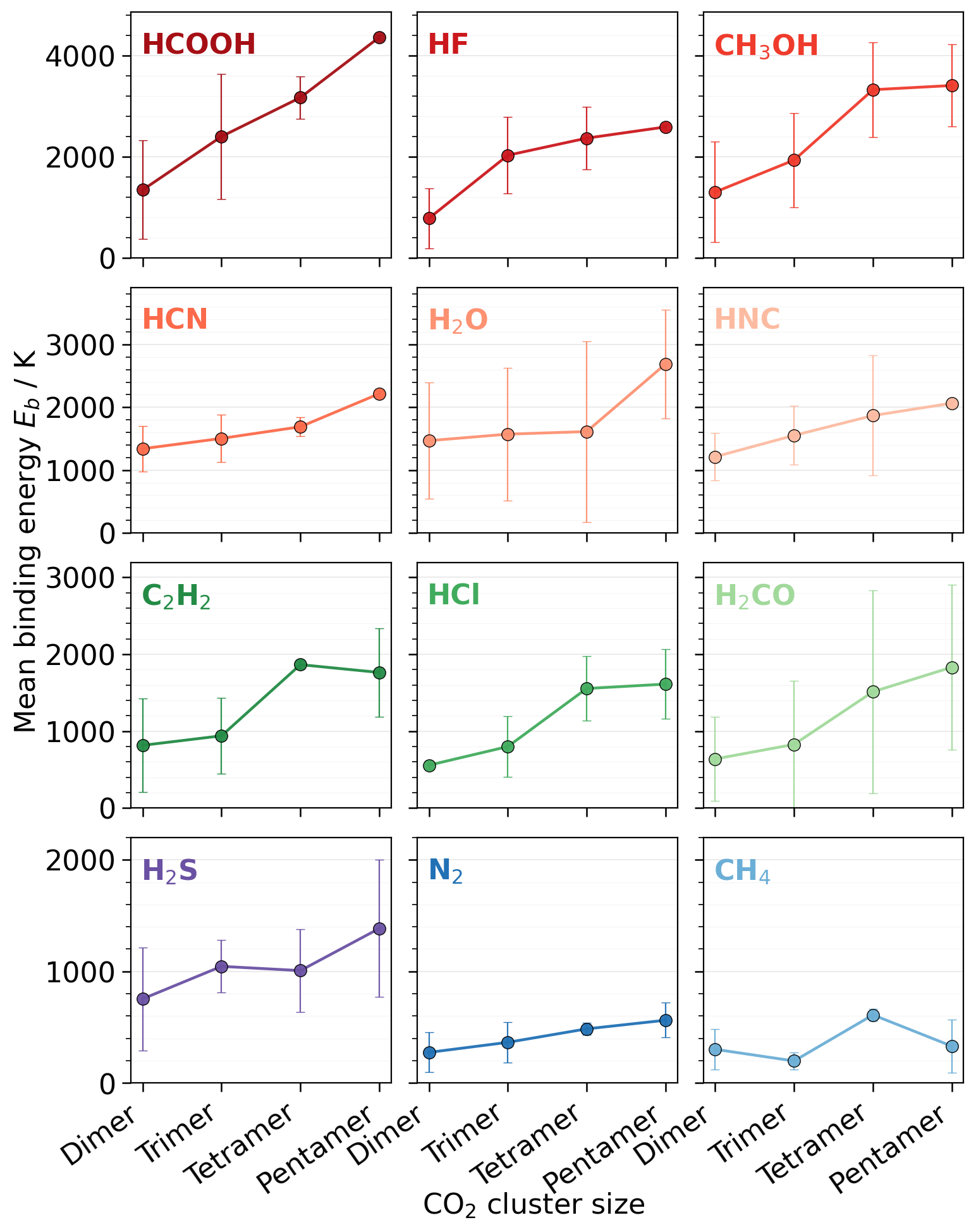}
    \caption{Mean binding energy $E_\mathrm{b}$ (K) versus CO$_2$ cluster size ($n = 2$--$5$) for the twelve adsorbates at $\omega$B97X-V/def2-TZVP, with error bars showing $\sigma_\mathrm{site}$ floored at the 53\,K benchmark MAE. Panels are colour-coded by binding-mode family: strong H-bond donors in warm reds, moderate binders in greens, weak and dispersion-bound species in blues, and H$_2$S in purple. H$_2$O shows the largest relative rise from tetramer to pentamer (+40\%), and HCOOH ($+27\%$) and HCN (+24\%) are also still rising steeply at $n = 5$. CH$_3$OH, sampled at four sites on both the tetramer and pentamer with a $+2\%$ step between them, shows the smallest tetramer-to-pentamer change of the strong donors. H$_2$S also takes a large step at the pentamer (+27\%), but as part of a non-monotonic size dependence rather than a smooth climb. Pentamer binding energy values for HF, HCN, HNC, and HCOOH derive from a single binding site ($\sigma$ floored at 53\,K), so the apparent flattening for HF and HNC is consistent with, but not a definitive demonstration of, the trend levelling off at $n = 5$.}
    \label{fig:be_convergence}
\end{figure*}
The three species behave differently at the tetramer-to-pentamer step. HCOOH keeps climbing steeply ($3172 \rightarrow 4364$\,K, $+27\%$; Figure~\ref{fig:be_convergence}), so its average BE is still rising at the cluster sizes reached here and the true bulk-ice value is expected to be higher still. CH$_3$OH, by contrast, grows by only $+2\%$ ($3327 \rightarrow 3410$\,K), with four sites at both the tetramer and pentamer, the smallest increase of the three strong donors across this step. HF grows by $+9\%$ ($2368 \rightarrow 2590$\,K), consistent with the trend levelling off, but resting on a single pentamer site, so we take the pentamer value as our best estimate for the desorption modelling in Section~\ref{sec:astro_implications}.

\begin{figure}
    \centering
    \includegraphics[width=3.1in, height=5in]{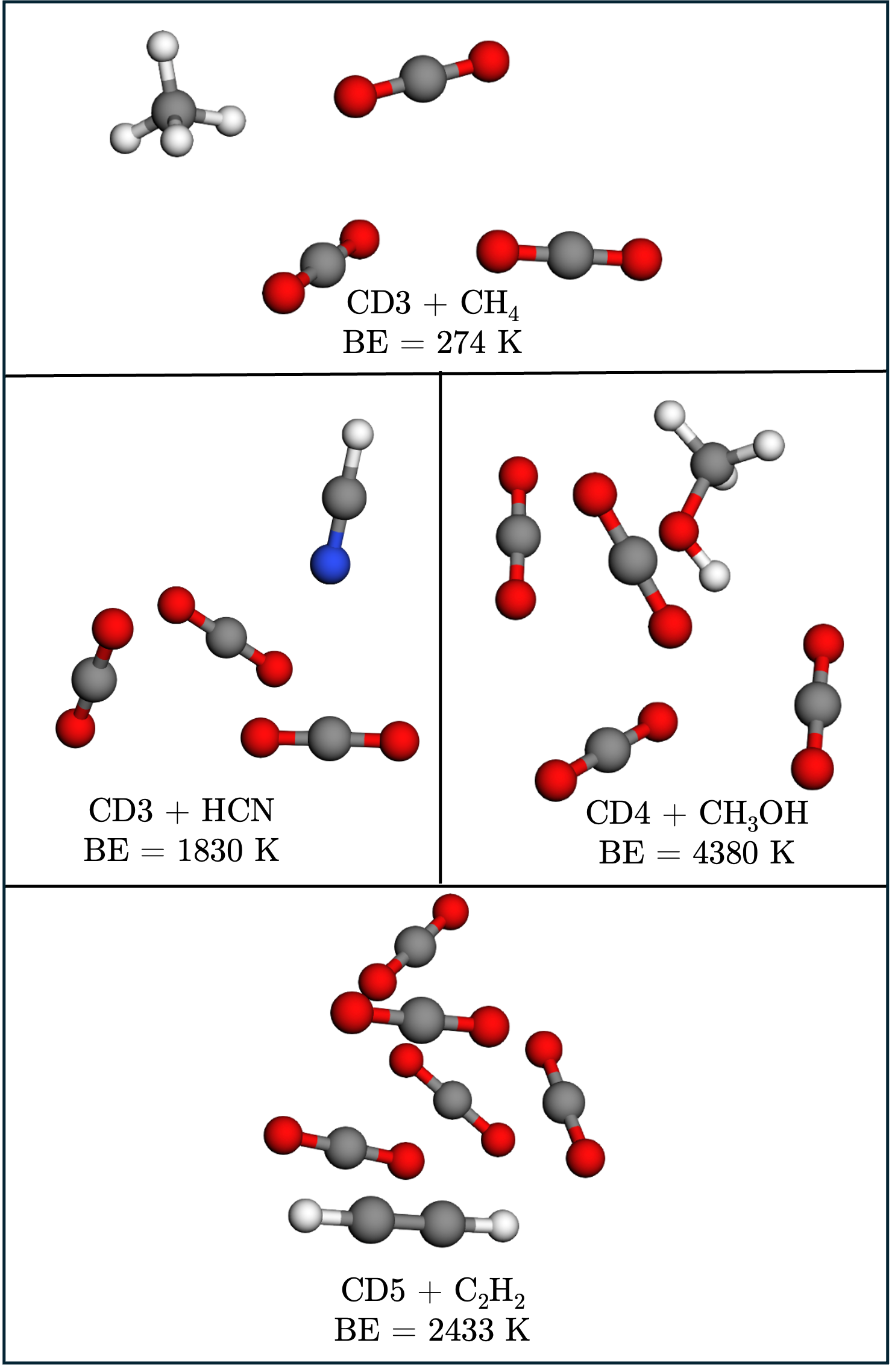}
    \caption{Representative binding sites illustrating the diversity of binding modes across the dataset. Top: CH$_4$ on CD$_3$-01 (274\,K): the deepest CH$_4$ trimer site and the weakest binding mode in the dataset, a pure dispersion contact with no polar donor or acceptor. Middle-left: HCN on CD$_3$-01 (1830\,K): the terminal N lone pair points toward an electron-poor CO$_2$ carbon. Middle-right: CH$_3$OH on CD$_4$-01 (4380\,K): the deepest O--H$\cdots$O=C donor in the dataset among the donors that lack an acidic proton. Bottom: C$_2$H$_2$ on CD$_5$-01 (2433\,K): the deepest C$_2$H$_2$ site, representative of the weakly acidic C$\equiv$C--H donor mode of the moderate-binder class.} 
    \label{fig:trends}
\end{figure}

HCN and HNC, the two cyanide isomers, both grow steadily with cluster size, from 1340\,K (HCN) and 1213\,K (HNC) at the dimer to 2216\,K and 2065\,K at the pentamer. Although the N--H bond of HNC is expected to be a better H-bond donor than the C--H bond of HCN, the two bind with similar strength throughout: HCN is higher at the dimer (by 127\,K) and the pentamer (by 151\,K), and HNC at the trimer (by 48\,K) and the tetramer (by 178\,K). HCN does not rely on its C--H bond, since it can also bind through its nitrogen lone pair, as at its deepest trimer site (Figure~\ref{fig:trends}). At the dimer, trimer, and tetramer the two distributions overlap within $\sigma_{\mathrm{site}}$, and both pentamer values come from a single binding site, so our data do not establish which isomer binds more strongly to CO$_2$. At the tetramer-to-pentamer step HNC grows by $9.5\%$, similar to HF, whereas HCN is still rising steeply ($+24\%$); with single-site pentamer values, neither trend can be confirmed without further sampling.

The mean BE of $\mathrm{H_2O}$ rises at every cluster size (1468, 1572, 1613, and 2689\,K from the dimer to the pentamer), but most of the increase comes at the last step, a $+40\%$ jump that is the largest relative tetramer-to-pentamer rise of any species. Water also has the largest site-to-site spread in the dataset, reaching $\sigma_{\mathrm{site}}=1437$\,K for the tetramer. Only three tetramer sites passed the site-selection criteria, so the tetramer mean, and with it the size of the final step, is poorly constrained. The optimised geometries show that H$_2$O binds through two complementary interactions: its oxygen approaches the electron-poor carbon of CO$_2$, while one of its hydrogens usually donates an H-bond to a CO$_2$ oxygen. The same two interactions are found in MP2 calculations of the isolated (CO$_2$)$_2$--H$_2$O complex \citep{Danten_2005}, and the BE of the 1:1 H$_2$O--CO$_2$ complex has been computed at correlated levels \citep{deLange_2011, Makarewicz_2010, Ramasami_2014}. However, CO$_2$ has no hydrogen with which to donate an H-bond back to water, so the reciprocal H-bonding available in water ice cannot form. The pentamer mean of 2689\,K is approximately $25\%$ lower than the 3592\,K obtained for water on water ice \citep{ahmad2026}. This difference is consistent with the more extensive H-bonding available on a water surface and may affect the distribution of $\mathrm{H_2O}$ and $\mathrm{CO_2}$ within mixed ice mantles (Section~\ref{sec:astro_implications}).
\subsubsection{Moderate binders}
\label{subsec:moderate_binders}
HCl, C$_2$H$_2$, H$_2$CO, and H$_2$S form an intermediate class, with pentamer BEs of $\sim 1400$--$1850$\,K. H$_2$CO binds most strongly of the four at the pentamer (1831\,K over $n = 18$ sites), growing from $\sim 640$\,K at the dimer through $\sim 830$\,K at the trimer as the larger clusters more often present the geometry its carbonyl requires. C$_2$H$_2$ follows closely (1763\,K over $n = 14$ sites). It grows from the dimer (815\,K) through the trimer (939\,K) to 1865\,K at the tetramer, where only a single site was retained, and its pentamer mean lies slightly below that value. Its C$\equiv$C--H$\cdots$O contact is stronger than acetylene's reputation as a purely dispersion-bound molecule would suggest, because the terminal C$\equiv$C--H proton is mildly acidic and can donate a weak H-bond. HCl reaches 1612\,K at the pentamer and has the largest binding-site population in the dataset, 25 sites on CD$_5$-01 ($\sigma_\mathrm{site} = 451$\,K); the H--Cl dipole acts as an effective H-bond donor to CO$_2$, and the wide spread of pentamer BEs reflects the many different positions and orientations HCl can take up on the irregular CD$_{5}$-01 surface. The H$_2$CO tetramer cell is poorly sampled under the site-selection criteria ($n = 2$, $\sigma_\mathrm{site} = 1321$\,K) and is not read as a distinct trend point.

The 18 $\mathrm{H_2CO}$ pentamer sites fall into two groups. At 11 sites, the carbonyl oxygen approaches an electron-poor CO$_2$ carbon at a distance of $2.80$--$2.88$\,\AA{}, giving a mean BE of 2543\,K. At the other seven sites, this distance increases to $3.8$--$5.2$\,\AA{}, and the molecule is bound mainly through weaker C--H$\cdots$O contacts, giving a mean BE of 713\,K. No intermediate O$\cdots$C distances occur in the sampled structures. These proportions describe the finite pentamer dataset and should not be interpreted as the expected populations in bulk $\mathrm{CO_2}$ ice.

The step from tetramer to pentamer is small for two of the four moderate binders. HCl grows by only $+4\%$, placing it with HF ($+9\%$), HNC ($+9.5\%$), and CH$_3$OH ($+2\%$) among the species whose mean BE has begun to level off by $n = 5$, and the C$_2$H$_2$ mean falls slightly ($-6\%$) from a tetramer value that rests on a single site. H$_2$CO ($+17\%$) and H$_2$S ($+27\%$) have not settled at this step, and neither has HCOOH, which is still climbing steeply ($+27\%$; discussed in Section~\ref{sec:H_bond_donors}). These percentage changes are computed from the mean BEs and indicate the size of the step between cluster sizes rather than being precisely determined; each mean carries the uncertainty given in Table~\ref{tab:final_BE}, and where the tetramer or pentamer value rests on a single site the change should be read as indicative only.

H$_2$S is the weakest of this class (1385\,K at the pentamer) despite carrying two hydrogens, because the weakly polarised S--H bond is a poor H-bond donor. Binding is therefore dominated by dispersion rather than directional donation to a CO$_2$ oxygen. The mean varies non-monotonically with cluster size (754, 1046, 1008, 1385\,K), dropping slightly at the tetramer before increasing by $27\%$ at the pentamer (Table~\ref{tab:final_BE}), and is therefore still changing at $n = 5$. The pentamer distribution is broad ($\sigma_{\mathrm{site}} = 615$\,K). The deepest site is also the most highly coordinated, consistent with dispersion strengthening in enclosed geometries through multiple simultaneous contacts. The mean therefore depends strongly on the sampling of such sites and may increase further for larger clusters.

The pentamer BEs for HCl, C$_2$H$_2$, H$_2$CO, and H$_2$O (Section~\ref{sec:H_bond_donors}) were computed on BLYP-D3BJ/def2-SVPD geometries, following the alternative procedure described in Section~\ref{subsec:rescue_protocol}, and therefore carry the additional methodological uncertainty of $\pm122$\,K quantified there, which is not included in $\sigma$ in Table~\ref{tab:final_BE}. For all four species the variation among binding sites is substantially larger than this uncertainty, so the error bars reported in Table~\ref{tab:final_BE} are not materially affected.
\subsubsection{Weak / dispersion-dominated binders}
\label{sec:weak_binders}
N$_2$ and CH$_4$ are the weakest binders. N$_2$ climbs from 275\,K at the dimer to 563\,K at the pentamer. Its $+14\%$ step from the tetramer leaves it rising at $n = 5$, although the tetramer value rests on a single site and the two cross-check functionals give a small decrease at this step (Section~\ref{sec:benchmark_main_body}). The pentamer distribution is comparatively narrow ($\sigma_\mathrm{site} =$ 156\,K over $n =$ 21 sites), because the dispersion contacts are similar wherever N$_2$ sits on the CD$_5$-01 surface. CH$_4$ shows no consistent trend with cluster size (302, 198, 611, and 331\,K): the high tetramer value comes from a single site, and the other three means agree within their $\sigma_\mathrm{site}$, consistent with non-directional dispersion binding that does not strengthen as more CO$_2$ molecules are added.

The pentamer BEs for N$_2$ and CH$_4$ were computed on BLYP-D3BJ/def2-SVPD geometries (Section~\ref{subsec:rescue_protocol}) and carry the additional 122\,K methodological uncertainty quantified there. For both species this is smaller than the tabulated $\sigma_\mathrm{site}$ (156 and 239\,K), so the site-to-site spread remains the dominant term. These two are nonetheless the hardest cases for the cross-check functionals: M06-HF/def2-TZVP has a large error relative to the CCSD(T)/CBS benchmark on the trimer for both (Appendix~\ref{sec:be_benchmark}) and leaves CH$_4$ unbound at the pentamer ($-28$\,K; Table~\ref{tab:be_three_functional}), so the functional spread is comparable to the BE itself. This is a known limitation of DFT for dispersion-bound systems rather than a flaw in the cluster set \citep{Grimme2010}, but the absolute BEs for N$_2$ and CH$_4$ should therefore be treated with caution.
\section{\textbf{Astrophysical Implications}} 
\label{sec:astro_implications}

\begin{table}[t]
\centering
\caption{Arithmetic mean binding energies in K and corresponding desorption peak temperatures, $T_{\mathrm{peak}}$, for adsorbates on the $\mathrm{CO_2}$ pentamer and amorphous solid water \citep{ahmad2026}.}
\label{tab:co2_asw_tpeak_combined}

\scriptsize
\setlength{\tabcolsep}{7pt}
\renewcommand{\arraystretch}{1.12}

\begin{tabular*}{\columnwidth}{@{\extracolsep{\fill}}lcc@{\hspace{1.2em}}cc@{}}
\hline\hline
& \multicolumn{2}{c}{$\mathrm{CO_2}$ pentamer}
& \multicolumn{2}{c}{$\mathrm{ASW_{12}}$} \\
\cline{2-3}\cline{4-5}
\noalign{\vskip 2pt}
Species
& $E_{\mathrm{b}}$ (K)
& $T_{\mathrm{peak}}$ (K)
& $E_{\mathrm{b}}$ (K)
& $T_{\mathrm{peak}}$ (K) \\
\hline
$\mathrm{CH_4}$   & 331  & 11  & 436  & 14  \\
$\mathrm{H_2CO}$  & 1831 & 57  & 2770 & 85  \\
$\mathrm{C_2H_2}$ & 1763 & 55  & 1921 & 59  \\
$\mathrm{H_2O}$   & 2689 & 82  & 3592 & 109 \\
$\mathrm{CH_3OH}$ & 3410 & 104 & 3909 & 119 \\
$\mathrm{HCOOH}$  & 4364 & 132 & 4881 & 147 \\
\hline
\end{tabular*}

\vspace{0.5em}

\begin{minipage}{\columnwidth}
\footnotesize
\textit{Note---}
The $\mathrm{CO_2}$ values are the arithmetic mean ZPVE-corrected binding energies calculated at $\omega$B97X-V/def2-TZVP for the retained sites on the pentamer cluster. The $\mathrm{ASW_{12}}$ values are arithmetic means over the binding sites reported by \citet{ahmad2026}. The corresponding $T_{\mathrm{peak}}$ values were obtained from the simulated desorption profiles.
\end{minipage}

\end{table}
\begin{figure*}
    \centering
    \includegraphics[width=7in, height=3.4in]{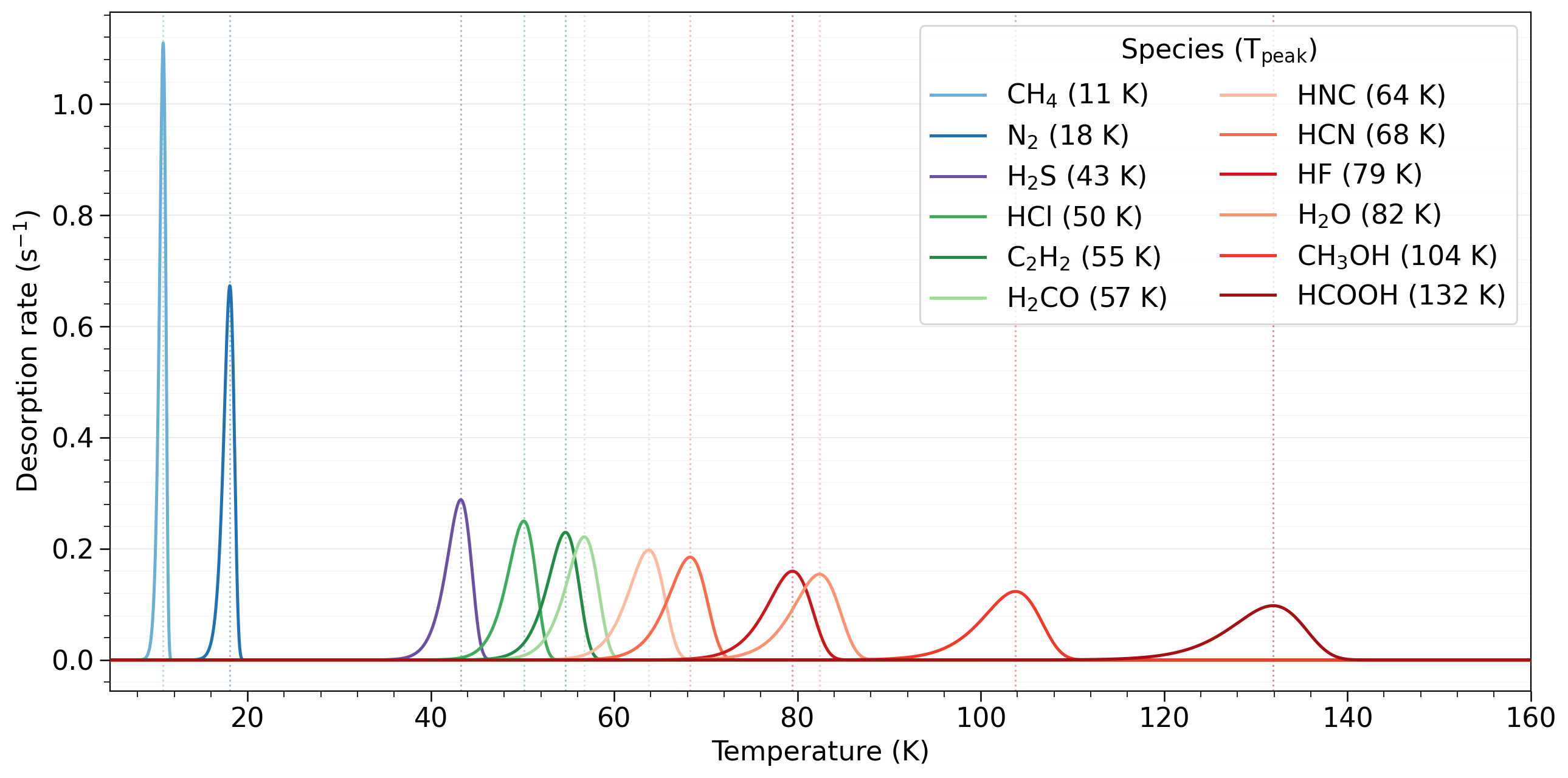}
    \caption{Simulated temperature-programmed desorption spectra for the twelve adsorbates on the CO$_2$ pentamer (CD$_5$-01), computed via first-order Polanyi-Wigner kinetics with a pre-exponential factor $\nu = 10^{12}$\,s$^{-1}$ and a laboratory-typical heating rate $\beta = 0.017$\,K\,s$^{-1}$ ($\approx 1$\,K\,min$^{-1}$). Each curve uses the cluster-mean ZPVE-corrected binding energy at $\omega$B97X-V/def2-TZVP (Table~\ref{tab:final_BE}). The temperature shown beside each species in the legend is its peak desorption temperature $T_\mathrm{peak}$, the temperature at which its desorption rate is highest, marked by the vertical dashed line under each curve.}
    \label{fig:tpd_cd5}
\end{figure*}
\begin{figure*}
    \centering
    \includegraphics[width=7in, height=3.4in]{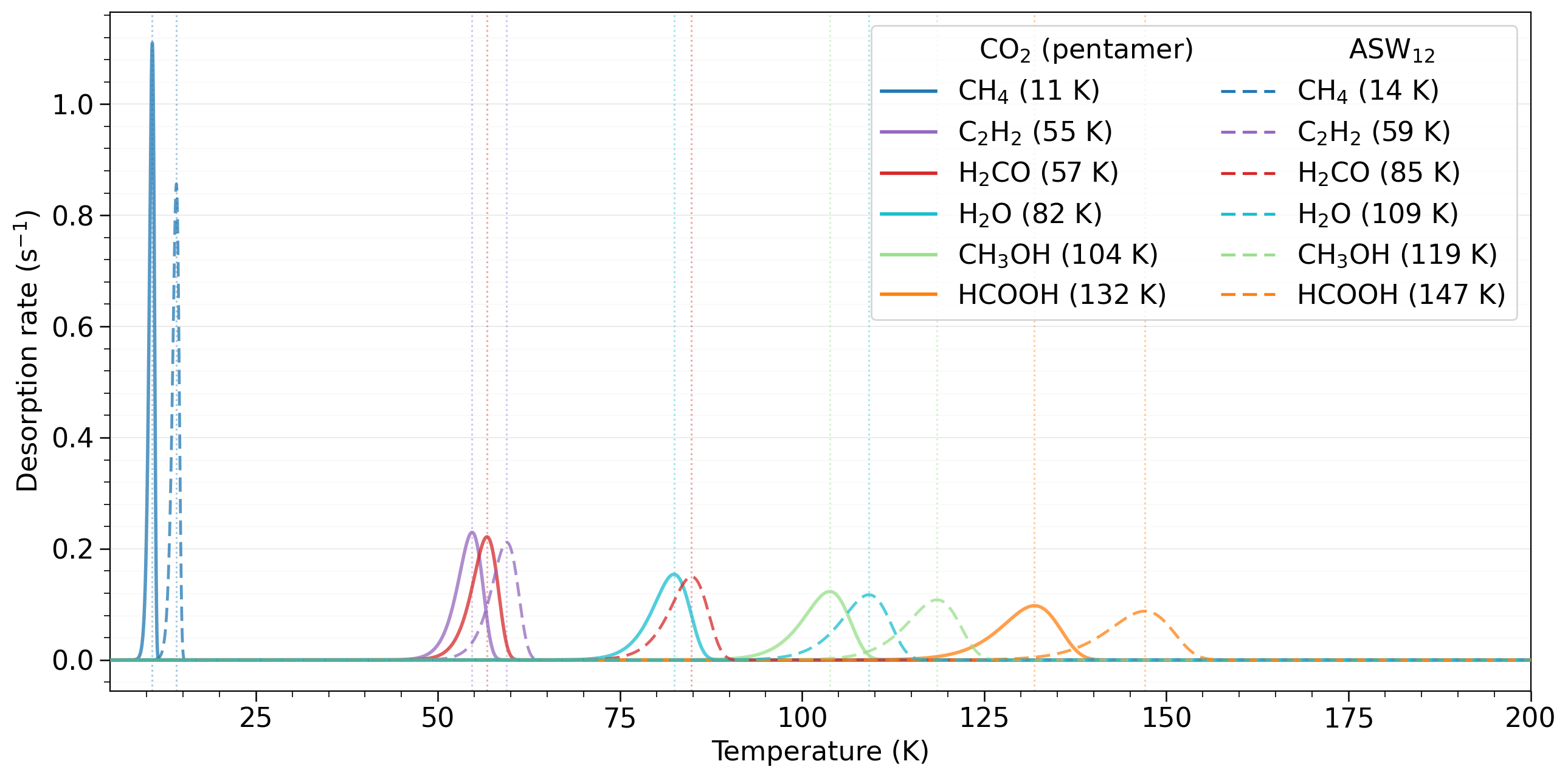}
    \caption{Simulated temperature-programmed desorption spectra of the six adsorbates studied in both this work and \citet{ahmad2026}, computed via first-order Polanyi-Wigner kinetics with a pre-exponential factor $\nu = 10^{12}$\,s$^{-1}$ and a laboratory-typical heating rate $\beta = 0.017$\,K\,s$^{-1}$ ($\approx 1$\,K\,min$^{-1}$). Solid curves use CO$_2$ pentamer binding energies at $\omega$B97X-V/def2-TZVP (Table~\ref{tab:final_BE}); dashed curves use the corresponding ASW$_{12}$ binding energy from \citet{ahmad2026}. The temperature shown beside each species in the legend is its peak desorption temperature $T_\mathrm{peak}$, the temperature at which its desorption rate is highest, marked by the vertical dashed line under each curve.}
    \label{fig:tpd_co2_vs_asw}

\end{figure*}
The BE distributions in Section~\ref{sec:BEs} translate directly into the temperatures at which each adsorbate thermally desorbs from a CO$_2$-rich surface, through the Polanyi--Wigner equation for first-order desorption kinetics:
\begin{equation}
    r = -\frac{d\theta}{dt} = \nu\,\theta^{n}\,\exp\!\left(-\frac{E_\mathrm{des}}{k_\mathrm{B} T}\right),
\end{equation}
where $r$ is the desorption rate, $\theta$ is the surface coverage, $\nu$ is the pre-exponential factor, $n$ is the desorption order, $E_\mathrm{des}$ is the desorption energy (here equal to the BE, $E_\mathrm{b}$), $k_\mathrm{B}$ is the Boltzmann constant, and $T$ is the temperature. We take monolayer desorption ($n = 1$); multilayer desorption ($n = 0$) is not considered, as we treat only molecule--surface interactions. Converting each BE to a peak desorption temperature $T_\mathrm{peak}$ under laboratory-typical conditions ($\nu = 10^{12}$\,s$^{-1}$ \citealt{Hasegawa_1992}, heating rate $\beta = 0.017$\,K\,s$^{-1} \approx 1$\,K\,min$^{-1}$ \citealt{Collings2004}) gives the spectra in Figure~\ref{fig:tpd_cd5} and the values in Table~\ref{tab:co2_asw_tpeak_combined}. Each curve uses a single BE per species, the mean over the retained pentamer sites (Table~\ref{tab:final_BE}); the BE distribution is not included. A single BE is commonly assumed in astrochemical models \citep{Cuppen_2017}. The peak temperatures span 11\,K (CH$_4$) to 132\,K (HCOOH), preserving the BE ordering of Section~\ref{sec:BEs}.

Laboratory BE measurements for molecules on CO$_2$ surfaces are scarce: most experimental BEs are determined for adsorption on water ice \citep[e.g.][]{Collings2004,He_2016}, and we are not aware of a systematic temperature-programmed desorption (TPD) dataset for stable molecules physisorbed on a pure CO$_2$ surface. The closest constraint is \citet{Luna_2008}, who found that pure CH$_4$ desorbs from solid CO$_2$ at 38\,K, while CH$_4$ trapped within a CO$_2$ matrix is retained to 50, 75, and up to 90\,K, released in stages as the CO$_2$ structure crystallises. These are entrapment temperatures set by the bulk matrix, not single-site surface binding, so they cannot calibrate our values directly, though they agree that CO$_2$ traps volatiles, holding them to higher temperatures than their pure-ice sublimation points.

Independently of any laboratory comparison, the $T_\mathrm{peak}$ values reported here are likely to be lower bounds on the bulk-ice values for the species whose mean BE still rises by more than 10\% at the pentamer (N$_2$, H$_2$S, H$_2$CO, HCN, HCOOH, and H$_2$O; Table~\ref{tab:final_BE}): a five-molecule cluster samples only a fraction of the binding geometries on the bulk CO$_2$ ice, and the deepest multi-coordinated sites that set the bulk $T_\mathrm{peak}$ are under-represented. The shortfall is likely to be largest for H$_2$O, HCOOH, H$_2$S, and HCN, whose means still rise at this step.\par
\begin{deluxetable*}{l@{\extracolsep{\fill}}rrrrrl}
\tabletypesize{\scriptsize}
\tablewidth{\textwidth}

\tablecaption{Binding energy statistics for the pentamer (CD$_5$-01) cluster.
The mean, standard deviation ($\sigma$), median, and range are reported in K. The number of binding sites is given by $n$. The trend column repeats the classification of Table~\ref{tab:final_BE}.
\label{tab:pentamer_statistics}}

\tablehead{
\colhead{Species} &
\colhead{Mean} &
\colhead{$\sigma$} &
\colhead{Median} &
\colhead{Range} &
\colhead{$n$} &
\colhead{Trend at $n=5$} \\
\colhead{} &
\colhead{(K)} &
\colhead{(K)} &
\colhead{(K)} &
\colhead{(K)} &
\colhead{} &
\colhead{}
}

\startdata
N$_2$      & 563  & 156                   & 543  & 269--753    & 21 & rising\tablenotemark{b} \\
CH$_4$     & 331  & 239                   & 316  & $-58$--754  & 15 & non-monotonic\tablenotemark{b} \\
HCl        & 1612 & 451                   & 1831 & 821--2657   & 25 & levelling \\
C$_2$H$_2$ & 1763 & 578                   & 1871 & 218--2433   & 14 & non-monotonic\tablenotemark{b} \\
H$_2$S     & 1385 & 615                   & 1248 & 875--2438   & 5  & non-monotonic \\
HF         & 2590 & 605\tablenotemark{a}  & \nodata & \nodata & 1 & levelling\tablenotemark{b} \\
HCN        & 2216 & 289\tablenotemark{a}  & \nodata & \nodata & 1 & rising steeply\tablenotemark{b} \\
H$_2$CO    & 1831 & 1073                  & 1943 & 409--3649   & 18 & rising \\
HCOOH      & 4364 & 749\tablenotemark{a}  & \nodata & \nodata & 1 & rising steeply\tablenotemark{b} \\
HNC        & 2065 & 786\tablenotemark{a}  & \nodata & \nodata & 1 & levelling\tablenotemark{b} \\
CH$_3$OH   & 3410 & 814                   & 3790 & 2193--3867  & 4  & levelling \\
H$_2$O     & 2689 & 864                   & 2665 & 1421--4345  & 23 & rising steeply
\enddata

\tablenotetext{a}{The uncertainty is taken from the pooled tetramer (CD$_4$-01) and pentamer (CD$_5$-01) distribution because only one pentamer binding site was obtained.}
\tablenotetext{b}{The trend assessment rests on a single site at the tetramer or pentamer (Table~\ref{tab:final_BE}) and should be treated as indicative.}

\end{deluxetable*}
For astrochemical modelling, we recommend the pentamer values in Table~\ref{tab:pentamer_statistics}, as these are based on the largest CO$_2$ clusters considered here and are therefore our closest approximation to an extended surface. The table reports the mean, standard deviation, median, and range across the available binding sites. For HNC, HCN, HF, and HCOOH, only one pentamer site passed the quality filter. For these species, we therefore estimate the spread from the combined tetramer and pentamer data. The pentamer values for N$_2$, CH$_4$, HCl, C$_2$H$_2$, H$_2$CO, and H$_2$O carry an additional methodological uncertainty of $\pm122$\,K (Section~\ref{subsec:rescue_validation}). We do not extrapolate to bulk CO$_2$ because the limited cluster-size range and non-monotonic trends leave the bulk values poorly constrained. Instead, the table repeats the trend at $n = 5$ from Table~\ref{tab:final_BE}. Values for species whose mean BE still rises by more than 10\% at the pentamer (N$_2$, H$_2$S, H$_2$CO, HCN, HCOOH, and H$_2$O) should be regarded as lower limits for an extended CO$_2$ surface. Larger, size-matched CO$_2$ clusters will be considered in future work to address this limitation (Section~\ref{sec:conc}).

The comparison with water ice (Table~\ref{tab:co2_asw_tpeak_combined}, Figure~\ref{fig:tpd_co2_vs_asw}), across the six species studied on both surfaces, shows that the difference between the two ices differs per species. This runs counter to how the two surfaces are treated in astrochemical models, which apply BEs referenced to water ice to all grains, whatever their composition \citep{Garrod_2011, Penteado_2017, Wakelam_2017}. The ASW$_{12}$ model is itself a small cluster, so its desorption temperatures are lower bounds too; the CO$_2$-to-water differences are meaningful only insofar as this bias is comparable on both ices. The contrast is largest for H$_2$CO and H$_2$O, which desorb at higher temperatures from water ice than from CO$_2$, with $\Delta T_{\mathrm{peak}}=+28$ and $+27$\,K, respectively. Formaldehyde shows the largest relative difference: its mean BE on water ice is approximately 1.5 times that on the CO$_2$ pentamer, while the corresponding ratio for water is approximately 1.3. Both differences have the same origin. Formaldehyde binds to CO$_2$ through its carbonyl oxygen, which donates electron density to an electron-poor CO$_2$ carbon; on water ice it can additionally accept an H-bond donated by a surface O--H, a contribution CO$_2$ cannot supply. Water is affected in the same way and more symmetrically: a CO$_2$ oxygen can accept an H-bond from water, but CO$_2$ has no hydrogen with which to donate one in return. Increasing the size of a pure-CO$_2$ cluster will not introduce this missing donor interaction, although it may still change the absolute BE.

The remaining species show the same sign but a smaller effect. CH$_3$OH and HCOOH peak 15\,K lower on the CO$_2$ pentamer than on water ice, and C$_2$H$_2$ only a few kelvin lower; all three bind a factor of $\sim 1.1$ more weakly on CO$_2$ (3410 versus 3909\,K for CH$_3$OH, 4364 versus 4881\,K for HCOOH, and 1763 versus 1921\,K for C$_2$H$_2$). That methanol is so little affected follows from the same donor--acceptor mechanism. Water in water ice is held almost entirely by H-bonds, donating and accepting in roughly equal measure, so on a CO$_2$ surface it loses the return bond it can no longer receive, and with it a large part of its binding. Methanol also donates and accepts, but its binding derives mainly from a single strong O--H donor bond, reinforced by dispersion from its methyl group, so the loss of the acceptor contact on CO$_2$ has little effect on its total BE.

Every species therefore binds more weakly on CO$_2$ than on water, but by a margin that ranges from a factor of $\sim 1.5$ for H$_2$CO, through $\sim 1.3$ for H$_2$O and CH$_4$, down to $\sim 1.1$ for C$_2$H$_2$, HCOOH and CH$_3$OH. These species are ordered above by $\Delta T_{\mathrm{peak}}$, the astrophysically relevant quantity, but the ratio orders them differently: CH$_4$ binds $\sim 1.3$ times more weakly on CO$_2$, nearly as large a proportional effect as water, yet shifts $T_{\mathrm{peak}}$ by only 3\,K because it is weakly bound on both surfaces. Astrochemical models do not capture this, because they use one set of BEs, referenced to water ice, for every grain surface \citep{Garrod_2011, Penteado_2017, Wakelam_2017}. Where the two surfaces bind a species almost equally, the difference is within our cluster-size uncertainty and cannot yet be confirmed; larger clusters are needed to resolve it.

These results inform how volatiles partition wherever CO$_2$ makes up a significant fraction of the grain-ice mantle, including the CO$_2$-rich regions seen in pre-stellar cores, protostellar envelopes, and protoplanetary-disk midplanes \citep{_berg_2011, Escribano_2013, Mcclure_2023, Sturm_2023, Brunken_2024, potapov2025, bergner2026}. In the absence of BEs for these species on CO$_2$, models of these regions use values measured on water ice \citep{Penteado_2017,Wakelam_2017}. This assumption does not hold uniformly. The magnitude of the CO$_2$-to-water difference varies species to species, considerably for some. The weak binding of water to CO$_2$ has little consequence where water ice dominates the surface, as it usually does; it matters only where CO$_2$ makes up most of the surface.

The clearest case is formaldehyde (H$_2$CO). H$_2$CO binds more weakly on CO$_2$ (1831\,K) than on water ice ($\sim$2770\,K), which lowers its thermal desorption temperature from $\sim$85\,K to $\sim$57\,K. This bears on the long-standing question of cold gas-phase H$_2$CO in disks. The two binding modes identified in Section~\ref{subsec:moderate_binders} are consistent with this asymmetry: on CO$_2$ the carbonyl either reaches a carbon or it does not, whereas the ASW$_{12}$ distribution is unimodal \citep{Bovolenta_2022, ahmad2026}, water offering a comparable arrangement at most sites. H$_2$CO is observed in the outer regions of disks as a ring beyond the CO snowline, in gas as cold as the $\sim$20\,K at which CO freezes onto grains, far below the temperature at which H$_2$CO would desorb from water ice \citep{Qi_2013, oberg_2017}. It is understood to form in situ by hydrogenation of CO within the CO-rich ice and to reach the gas by non-thermal desorption \citep{Qi_2013, oberg_2017}. Our result modifies the thermal part of this picture. CO$_2$ is concentrated in the same CO-rich ice in which H$_2$CO forms, both because it forms there and because CO evaporates and leaves CO$_2$ behind; if the ice retains its layered structure, the H$_2$CO is left in contact with CO$_2$ rather than water once the CO has been released. Bound to CO$_2$, it would desorb thermally at $\sim$57\,K rather than $\sim$85\,K, so its snowline shifts outward to a colder radius and the region over which it is thermally in the gas is correspondingly larger. These are laboratory desorption temperatures; under longer timescales and lower pressures of a real disk, desorption occurs at lower temperatures still, which could place the CO$_2$-associated H$_2$CO snowline as low as $\sim$45\,K. This does not by itself account for the H$_2$CO seen near the CO snow line at $\sim$20\,K, which is colder again and requires non-thermal desorption. What it does is move the boundary between thermal and non-thermal desorption outward, predicting a wider thermally desorbed H$_2$CO reservoir than a water-ice BE would.

This picture is kinetic rather than thermodynamic. $\mathrm{H_2CO}$ ends up with $\mathrm{CO_2}$ not because that is its most stable location but because it forms there and cannot diffuse away before the CO leaves. H$_2$CO diffuses slowly through ice, and once buried in an ice mantle its desorption is held back by the need to reach the surface first, as measured in water ice by \citet{Mispealer_2013}. The two processes compete, since $\mathrm{CO_2}$ segregation and $\mathrm{H_2CO}$ desorption occur over similar temperature ranges, so whether the H$_2$CO is still in a CO$_2$-rich ice environment when it desorbs is the open question.

The same asymmetry explains why mixed CO$_2$--H$_2$O ice separates into pure CO$_2$ regions. Water binds more weakly to a CO$_2$ surface than to water ice, so once the ice is warm enough to rearrange, the two pull apart. This is what experiments find. \citet{He_2017} measured the CO$_2$--CO$_2$ BE to be $2415 \pm 20$\,K and the CO$_2$--water BE $2250 \pm 20$\,K, and observed directly that CO$_2$ does not wet a water-ice surface but gathers into clusters. Our computed CO$_2$--H$_2$O (CO$_2$ pentamer) BE of 2689\,K is $25\%$ below the 3592\,K we obtain for water on water ice, confirming the asymmetry from the opposite side and providing the first-principles reason for it: water cannot receive the return H-bond that binds it in water ice, whereas CO$_2$ binds CO$_2$ without needing one. The picture fits the pure CO$_2$ ice seen toward protostars by both Spitzer \citep{Knez_2005, Pontoppidan_2008, Escribano_2013} and JWST \citep{Mcclure_2023, Brunken_2024}. A direct comparison with matched water-ice and mixed ice clusters is left to future work.
\section{\textbf{Conclusions}} 
\label{sec:conc}

We have computed ab initio BEs for twelve astrophysically relevant molecules on CO$_2$ ice clusters of size $n=2$--$5$, using the BEEP platform with a production functional ($\omega$B97X-V/def2-TZVP) calibrated against a CCSD(T)/CBS reference. The resulting set of BEs for astrophysically relevant molecules on CO$_2$ clusters identifies the mechanism that determines how strongly each species binds: whether the adsorbate requires the surface to donate a H-bond back to it. This mechanism is local and largely independent of cluster size, so the trends are expected to carry over to bulk ice even where the absolute BEs may not.

Tracking the mean BE with cluster size separates the species whose BE has begun to level off from those still rising. CH$_3$OH shows the smallest change of the strong donors, only $+2\%$ between the tetramer and pentamer, while HCOOH climbs steeply and monotonically at this step ($+27\%$), so its bulk-ice value lies higher than the cluster sizes here can capture. H$_2$O rises more steeply still ($+40\%$), with most of its increase coming at the last step, although its tetramer mean rests on only three sites. H$_2$S takes a similar step to HCOOH ($+27\%$) as part of a non-monotonic size dependence rather than a smooth climb.

Converting the BEs to desorption temperatures and comparing with water ice results from \citet{ahmad2026} shows that every species binds more weakly on CO$_2$, but by a margin that varies substantially from species to species, contrary to the implicit assumption in astrochemical models that use water-ice values for every grain surface. It ranges from $\sim 1.5$ for formaldehyde to $\sim 1.1$ for acetylene, formic acid, and methanol. This variation follows from the binding mechanism: the species most affected are those whose binding on water ice includes an H-bond donated by the surface, a contribution CO$_2$ cannot make and that adding more CO$_2$ cannot remedy, whereas molecules that bind through their own donating groups are far less affected. Formaldehyde shows the largest effect, and its lower desorption temperature on CO$_2$ may bear on the cold gas-phase formaldehyde seen in disks. The same asymmetry provides a first-principles basis for the segregation of mixed H$_2$O--CO$_2$ ice into pure CO$_2$ regions, consistent with both laboratory measurements and the pure CO$_2$ ice observed toward protostars.

The natural next step is to compare the two ices using more closely matched surfaces. The present comparison sets our five-molecule CO$_2$ pentamer against the twelve-molecule ASW$_{12}$ water-ice model \citep{ahmad2026}, so part of every CO$_2$-to-water difference reflects the difference in surface size rather than surface chemistry. We are therefore extending this work to a twelve-molecule CO$_2$ cluster, matched to ASW$_{12}$ in the number of surface molecules. Low-energy (CO$_2$)$_n$ structures are available at the MP2 level up to $n = 13$ \citep{Yeole_2012}, and a single H$_2$O molecule is surrounded by seven CO$_2$ molecules in its first solvation shell \citep{Thanthiriwatte_2012}, so a twelve-molecule cluster is large enough to accommodate it. Pure CO$_2$ clusters nonetheless reach the bulk cubic structure only at around one hundred molecules \citep{Torchet_1984}; the twelve-molecule cluster is chosen to match ASW$_{12}$ rather than to approach the bulk limit. Neither cluster reaches the bulk-ice limit, and matching the molecule count does not make the substrates equivalent in every respect, because CO$_2$ and H$_2$O differ in molecular size and in the number of contacts they offer an adsorbate. It does, however, remove the largest source of mismatch in the present comparison, where the CO$_2$ surfaces contain fewer than half as many molecules as the water surfaces. That study will turn the mechanistic picture established here into the quantitative comparison that the present size mismatch does not yet allow.

A further step, prompted by the JWST finding that CO$_2$ is often mixed with water rather than pure \citep{Mcclure_2023, bergner2024, bergner2026}, is to compute BEs on mixed H$_2$O--CO$_2$ ices. There a molecule may bind against water, against CO$_2$, or at a boundary between them, and its BE will fall between the two pure-ice values established here and in our earlier work. These mixed, or \textit{'dirty'}, ices are the environment much of the observed CO$_2$ actually occupies \citep{Pontoppidan_2008, Mcclure_2023}, and describing them is what a complete astrochemical treatment of CO$_2$-rich grains ultimately requires.
\begin{acknowledgments}
A.A. acknowledges support from UK Research and Innovation (grant numbers MR/T040726/1 and MR/Z00029X/1). C.W. acknowledges financial support from the Science and Technology Facilities Council and UK Research and Innovation (grant numbers ST/X001016/1, MR/T040726/1 and MR/Z00029X/1). S.V-G. acknowledges financial support from the VRID research grant 2022000507INV. We thank A. Lala, who contributed to the theoretical calculations of the dimer and trimers during his master's project at the University of Leeds.
\end{acknowledgments}
\bibliography{Papers}{}
\bibliographystyle{aasjournalv7}
\clearpage
\FloatBarrier
\onecolumngrid
\appendix

This appendix presents in full the two validation exercises summarised in Section~\ref{sec:benchmark_main_body}, together with the supporting tables. Table~\ref{tab:rescue_protocol_validation} reports the control study validating the use of BLYP-D3BJ for the unconverged pentamer sites (Section~\ref{subsec:rescue_protocol}). Appendix~\ref{sec:geometry_consistency} describes the geometry consistency check against the MP2/def2-TZVP reference and its results, supported by the full per-method RMSD ranking (Table~\ref{tab:geom_full}), the per-adsorbate heavy-atom Kabsch RMSDs (Table~\ref{tab:rmsd_methods}), and the basin analysis for the two adsorbates that populate more than one low-lying minimum (Table~\ref{tab:outlier_analysis}). Appendix~\ref{sec:be_benchmark} describes the 114-functional binding-energy benchmark against the CCSD(T)/CBS reference and its results (Table~\ref{tab:be_bench_full}), and Table~\ref{tab:be_three_functional} reports the mean BEs at the three production functionals across the CD2--CD5 clusters.
\begin{table*}[ht]
\centering
\scriptsize
\caption{Validation of the rescue protocol using 24 sites for which the MPWB1K-D3BJ/def2-SVPD refinement had converged. Each site was reoptimised at BLYP-D3BJ/def2-SVPD, and the CP-corrected BE and $\Delta$ZPVE were recalculated on the resulting geometry. RMSD is the heavy-atom Kabsch deviation between the geometries obtained with the two methods. The reported differences are defined as the BLYP value minus the MPWB1K value, with $\Delta E_{\mathrm{b}}=\Delta\mathrm{BE}+\Delta\mathrm{ZPVE}$; a positive value therefore indicates stronger binding at the rescue level. Values of $\Delta$ZPVE are unavailable for HCN/CD$_4$-01 and two N$_2$/CD$_4$-01 sites because their BLYP Hessian calculations did not converge. Three BLYP optimisations reached a different local minimum, identified by an RMSD greater than $0.3$\,\AA{}. We include these sites in the mean because comparable changes in geometry may occur among the rescued sites, for which no converged MPWB1K geometry is available. For HCl/CD$_2$-01, an RMSD of $1.2$\,\AA{} corresponds to a relatively small $|\Delta E_{\mathrm{b}}|$ of $60$\,K. The median $|\Delta E_{\mathrm{b}}|$ across the 21 complete comparisons is $58$\,K.}
\label{tab:rescue_protocol_validation}

\begin{tabular*}{\textwidth}{@{\extracolsep{\fill}}llrrrr}
\hline\hline
Site & Cluster & RMSD & $\Delta$BE & $\Delta$ZPVE &
$\Delta E_{\mathrm{b}}$ \\
     &         & (\AA) & (K) & (K) & (K) \\
\hline
HCl          & CD$_2$-01 & 1.199 &  111 & $-171$ &  $-60$ \\
CH$_4$       & CD$_3$-01 & 0.097 &    2 &  $-61$ &  $-59$ \\
H$_2$O       & CD$_3$-01 & 0.258 & $-134$ &  $-13$ & $-148$ \\
HCOOH        & CD$_3$-02 & 0.056 &    4 &   173 &   177 \\
C$_2$H$_2$   & CD$_4$-01 & 0.113 &  125 &  $-74$ &    51 \\
CH$_4$       & CD$_4$-01 & 0.076 &    5 &  $-99$ &  $-93$ \\
CH$_4$       & CD$_4$-01 & 0.097 &   86 &  $-64$ &    22 \\
H$_2$CO      & CD$_4$-01 & 0.224 &  219 & $-225$ &   $-6$ \\
H$_2$CO      & CD$_4$-01 & 0.094 &   12 &    51 &    62 \\
H$_2$CO      & CD$_4$-01 & 0.079 &    8 &  $-63$ &  $-55$ \\
H$_2$CO      & CD$_4$-01 & 0.076 &    0 &  $-21$ &  $-21$ \\
H$_2$CO      & CD$_4$-01 & 0.384 &  463 & $-132$ &   332 \\
H$_2$O       & CD$_4$-01 & 1.134 & 1100 & $-162$ &   938 \\
H$_2$O       & CD$_4$-01 & 0.141 &   42 &   128 &   169 \\
H$_2$O       & CD$_4$-01 & 0.073 &    1 &    53 &    54 \\
HCl          & CD$_4$-01 & 0.072 &  $-14$ &  $-36$ &  $-50$ \\
HCl          & CD$_4$-01 & 0.114 &   52 &  $-28$ &    24 \\
HCl          & CD$_4$-01 & 0.080 &    7 &  $-63$ &  $-56$ \\
HCl          & CD$_4$-01 & 0.081 &    4 &  $-62$ &  $-58$ \\
HCl          & CD$_4$-01 & 0.108 &    3 &  $-41$ &  $-38$ \\
HCN          & CD$_4$-01 & 0.103 &   $-9$ & \nodata & \nodata \\
N$_2$        & CD$_4$-01 & 0.153 &   14 & \nodata & \nodata \\
N$_2$        & CD$_4$-01 & 0.129 &   $-2$ &  $-91$ &  $-93$ \\
N$_2$        & CD$_4$-01 & 0.077 &    1 & \nodata & \nodata \\
\hline
\multicolumn{5}{l}{Mean $|\Delta E_{\mathrm{b}}|$, 21 sites} & 122 \\
\multicolumn{5}{l}{Mean $|\Delta E_{\mathrm{b}}|$, excluding the three geometry changes} & 69 \\
\hline
\end{tabular*}
\end{table*}
\newpage
\onecolumngrid
\section{Benchmarking}
\label{sec:benchmarking}
\subsection{Geometry consistency check}
\label{sec:geometry_consistency_method}
\label{sec:geometry_consistency}

To check our geometry optimisation method (MPWB1K-D3BJ/def2-SVPD \citep{Zhao2004, Rappoport2010}; Section~\ref{subsec:beep_workflow}) against higher-level references, and to map out the shape of the energy surface for each complex, we re-optimised each of the twelve adsorbate--CO$_2$ monomer (CD$_1$-01) complexes from the same starting geometry at 25 levels of theory. These were: MP2 with the def2-TZVP basis \citep{moller1934,Weigend2005}, which serves as the wavefunction reference; three composite methods designed for fast, reasonable geometries (HF-3c/MINIX, PBEh-3c/def2-mSVP \citep{grimme2015}, and r$^2$SCAN-3c/def2-mTZVPP \citep{grimme2021}); and twenty-one density functionals with the def2-TZVP basis, drawn from all the major families and including MPWB1K-D3BJ/def2-SVPD. The complete list is given in Table~\ref{tab:geom_full}. Agreement is measured as the heavy-atom residual RMSD after optimal alignment: the average distance between corresponding non-hydrogen atoms once the two structures are rotated to align as closely as possible using the Kabsch algorithm.

Of the 300 optimisations attempted (25 methods $\times$ 12 adsorbates), 292 completed without error, and 247 of those reached the optimiser's convergence criteria. Two combinations, CH$_3$OH with $\omega$B97M-V and HCOOH with $\omega$B97X-V, failed for numerical rather than physical reasons. All statistics reported below are computed over converged geometries only; unconverged optimisations are excluded rather than being treated as predicted structures.

Some molecules can adsorb on the CO$_2$ in more than one stable position, each corresponding to a separate dip in the energy landscape, or basin. Different methods sometimes settle into different basins. Where the geometries from different methods disagreed by more than 1.0\,\AA{}, we organised them into these positions and measured how far the molecule sat from the centre of the CO$_2$. This distance shows whether the molecule is bound close to the surface or has drifted away and is barely held.

Across the 12-adsorbate test set, the MPWB1K-D3BJ/def2-SVPD geometries agree with MP2 to within 0.1\,\AA{} for 9 of the 10 adsorbates whose optimisation converged, with a median RMSD of 0.034\,\AA{} and a mean of 0.149\,\AA{}; the mean is inflated by a single outlier (HCOOH, discussed below). Among the five methods compared per adsorbate in Table~\ref{tab:rmsd_methods}, the range-separated $\omega$B97X-D/def2-TZVP geometries agree most closely with the reference (mean 0.029\,\AA{}, max 0.101\,\AA{}), consistent with its combination of a triple-zeta basis and an explicit dispersion treatment; across the fuller method set of Table~\ref{tab:geom_full}, $\omega$B97X-V/def2-TZVP and PBE-D3BJ/def2-TZVP agree more closely still (mean 0.010\,\AA{} each). The composite methods HF-3c/MINIX, r$^2$SCAN-3c/def2-mTZVPP, and PBEh-3c/def2-mSVP yield mean RMSDs of 0.135, 0.189, and 0.287\,\AA{}, respectively. Geometry-optimisation convergence rates were 12/12 (HF-3c), 11/12 (PBEh-3c), 10/12 (MPWB1K-D3BJ and $\omega$B97X-D), and 9/12 (r$^2$SCAN-3c); the MP2 reference converged for all twelve adsorbates. Per-adsorbate values are given in Table~\ref{tab:rmsd_methods}.

Two adsorbates (H$_2$CO and HCOOH) show pairwise method-to-method RMSDs exceeding 1\,\AA{} and were examined individually (Table~\ref{tab:outlier_analysis}). Combined with the adsorbate--CO$_2$ centroid distance, the pairwise RMSDs reveal that these outliers reflect the existence of multiple low-lying minima on the CD$_1$-01 + adsorbate PES, rather than a systematic failure of any one method. For H$_2$CO, four of six methods, including both the production MPWB1K-D3BJ/def2-SVPD and the MP2/def2-TZVP reference, locate a longer-range minimum at $d_{\mathrm{centroid}} \approx 4.5$\,\AA{} (MPWB1K--MP2 RMSD $= 0.091$\,\AA{}), while PBEh-3c and r$^2$SCAN-3c locate a closer-bound minimum at $d_{\mathrm{centroid}} \approx 3.5$\,\AA{}. The production geometry is therefore consistent with the wavefunction reference for this adsorbate.

For HCOOH, two distinct basins are populated: HF-3c, PBEh-3c, and the production MPWB1K-D3BJ/def2-SVPD cluster at $d_{\mathrm{centroid}} \approx 5.0$--$5.3$\,\AA{}, while r$^2$SCAN-3c, $\omega$B97X-D, and the MP2 reference agree closely at $\approx 3.83$--$3.90$\,\AA{} (RMSD $\le 0.040$\,\AA{} from MP2). The production method therefore locates a different shallow monomer orientation from the reference, whereas the two methods combining a triple-zeta-quality basis with an explicit dispersion treatment reproduce the MP2 minimum. This disagreement is confined to the CD$_1$-01 monomer, where HCOOH makes only a single contact and the competing orientations are near-degenerate, so the choice of lowest minimum is sensitive to the method. It does not propagate to the $n = 2$--$5$ clusters that underlie our reported BEs: from the dimer upward, the --COOH group bridges two CO$_2$ units, donating through O--H$\cdots$O=C while the carbonyl oxygen engages the electrophilic carbon of a neighbouring CO$_2$ (Section~\ref{sec:BEs}, Figure~\ref{fig:hcooh}), a doubly anchored geometry that is unambiguous across methods. The monomer is in any case excluded from the binding-site analysis and serves only as the consistency-check substrate. No single method is uniformly reliable across the two outlier cases: r$^2$SCAN-3c/def2-mTZVPP reproduces the MP2 minimum for HCOOH but selects the closer-bound basin for H$_2$CO, while MPWB1K-D3BJ/def2-SVPD does the reverse. This reinforces that the disagreements reflect near-degenerate minima on the monomer PES rather than a systematic deficiency of any one method.

All five methods agree that N$_2$ is essentially unbound on the CO$_2$ monomer, converging to a near-dissociated configuration at $\sim 6.3$\,\AA{} with a maximum deviation of 0.011\,\AA{} between methods. This reflects the small number of contacts a single CO$_2$ offers rather than any deficiency of the methods, and is not seen on the larger clusters, where N$_2$ interacts with several CO$_2$ molecules at once (e.g.\ 275\,K on the dimer; Table~\ref{tab:final_BE}).
\begin{table*}[t]
\centering
\caption{Full per-method summary of the geometry consistency check.}
\label{tab:geom_full}

\scriptsize
\setlength{\tabcolsep}{5pt}
\renewcommand{\arraystretch}{1.12}

\begin{tabular}{cllcrrr}
\hline\hline
Rank & Method / basis & Category & $N_\mathrm{conv}$ & Mean & Median & Max \\
     &                &          &                   & (\AA) & (\AA)  & (\AA) \\
\hline
--- & \textit{MP2/def2-TZVP} (ref.) & \textit{wavefunction} & 12 & 0.000 & 0.000 & 0.000 \\
1 & $\omega$B97X-V/def2-TZVP & DFT & 11 & 0.010 & 0.008 & 0.018 \\
2 & PBE-D3BJ/def2-TZVP & DFT & 10 & 0.010 & 0.009 & 0.020 \\
3 & PBE0-D3BJ/def2-TZVP & DFT & 11 & 0.013 & 0.013 & 0.021 \\
4 & B3LYP-D3BJ/def2-TZVP & DFT & 11 & 0.014 & 0.016 & 0.021 \\
5 & $\omega$B97M-V/def2-TZVP & DFT & 11 & 0.021 & 0.018 & 0.052 \\
6 & B97M-V/def2-TZVP & DFT & 11 & 0.021 & 0.019 & 0.042 \\
7 & BLYP-D3BJ/def2-TZVP & DFT & 11 & 0.023 & 0.009 & 0.130 \\
8 & M05-2X/def2-TZVP & DFT & 9 & 0.023 & 0.019 & 0.043 \\
9 & PWB6K/def2-TZVP & DFT & 9 & 0.025 & 0.018 & 0.068 \\
10 & TPSSH-D3BJ/def2-TZVP & DFT & 10 & 0.029 & 0.029 & 0.057 \\
11 & $\omega$B97X-D/def2-TZVP & DFT & 10 & 0.029 & 0.016 & 0.101 \\
12 & M06-2X/def2-TZVP & DFT & 10 & 0.048 & 0.050 & 0.096 \\
13 & SCAN/def2-TZVP$^{\dagger}$ & DFT & 3 & 0.057 & 0.059 & 0.063 \\
14 & B3LYP/def2-TZVP & DFT & 10 & 0.069 & 0.040 & 0.288 \\
15 & PBE0/def2-TZVP & DFT & 12 & 0.134 & 0.030 & 1.208 \\
16 & HF-3c/MINIX & 3c composite & 12 & 0.135 & 0.039 & 1.175 \\
\textbf{17} & \textbf{MPWB1K-D3BJ/def2-SVPD} & \textbf{DFT} & \textbf{10} & \textbf{0.149} & \textbf{0.034} & \textbf{1.174} \\
18 & PBE/def2-TZVP & DFT & 11 & 0.152 & 0.038 & 1.224 \\
19 & M06-L/def2-TZVP & DFT & 9 & 0.179 & 0.031 & 1.304 \\
20 & r$^2$SCAN-3c/def2-mTZVPP & 3c composite & 9 & 0.189 & 0.040 & 1.403 \\
21 & TPSSH/def2-TZVP & DFT & 10 & 0.202 & 0.071 & 1.250 \\
22 & TPSS/def2-TZVP & DFT & 11 & 0.203 & 0.112 & 1.263 \\
23 & PBEh-3c/def2-mSVP & 3c composite & 11 & 0.287 & 0.070 & 1.397 \\
24 & SCAN-D3BJ/def2-TZVP$^{\dagger}$ & DFT & 3 & 0.478 & 0.066 & 1.314 \\
\hline
\end{tabular}

\vspace{0.5em}

\begin{minipage}{0.92\textwidth}
\footnotesize
\textit{Note.} Heavy-atom Kabsch RMSDs are reported for optimised CD-01+adsorbate complexes relative to the MP2/def2-TZVP reference across 12 adsorbates. Methods are sorted by mean RMSD over converged sites only; $N_\mathrm{conv}$ is the number of converged optimisations contributing to each method's statistics out of 12 attempts. The production geometry optimisation level, MPWB1K-D3BJ/def2-SVPD, is highlighted in bold. SCAN and SCAN-D3BJ converged for only 3 of 12 sites, so their statistics rest on a small sample and should be interpreted with caution.
\end{minipage}

\end{table*}
\begin{deluxetable*}{lccccc}
\tabletypesize{\scriptsize}
\tablecaption{Heavy-atom Kabsch RMSD values (\AA) relative to the MP2/def2-TZVP reference geometry.\label{tab:rmsd_methods}}
\tablewidth{0pt}
\tablehead{
\colhead{Adsorbate} &
\colhead{HF-3c/MINIX} &
\colhead{PBEh-3c/def2-mSVP} &
\colhead{r$^2$SCAN-3c/def2-mTZVPP} &
\colhead{MPWB1K-D3BJ/def2-SVPD} &
\colhead{$\omega$B97X-D/def2-TZVP}
}
\startdata
C$_2$H$_2$ & 0.041 & 0.011 & 0.005 & 0.016 & 0.040 \\
CH$_3$OH   & 0.076 & \nodata & \nodata & 0.038 & \nodata \\
CH$_4$     & 0.038 & 0.070 & 0.032 & 0.021 & 0.012 \\
H$_2$CO    & 0.054 & \textbf{1.397} & \textbf{1.403} & 0.091 & 0.101 \\
H$_2$O     & 0.065 & 0.012 & \nodata & 0.014 & 0.012 \\
H$_2$S     & 0.028 & 0.052 & \nodata & \nodata & \nodata \\
HCN        & 0.085 & 0.034 & 0.038 & 0.044 & 0.046 \\
HCOOH      & \textbf{1.176} & \textbf{1.258} & 0.040 & \textbf{1.174} & 0.030 \\
HCl        & 0.004 & 0.124 & 0.080 & 0.024 & 0.019 \\
HF         & 0.020 & 0.094 & 0.047 & 0.031 & 0.013 \\
HNC        & 0.023 & 0.092 & 0.046 & 0.037 & 0.008 \\
N$_2$      & 0.007 & 0.010 & 0.007 & \nodata & 0.011 \\
\hline
$N_\mathrm{conv}$ & 12 & 11 & 9 & 10 & 10 \\
Mean       & 0.135 & 0.287 & 0.189 & 0.149 & 0.029 \\
Median     & 0.039 & 0.070 & 0.040 & 0.034 & 0.016 \\
Max        & 1.176 & 1.397 & 1.403 & 1.174 & 0.101 \\
\enddata
\tablecomments{Bold values indicate outliers, defined here as RMSD $> 0.5$~\AA{}, corresponding to a distinct local minimum relative to the MP2/def2-TZVP reference geometry. Entries marked \nodata{} are optimisations that did not reach the convergence criteria and are excluded from the statistics; $N_\mathrm{conv}$ is the number of converged optimisations contributing to each column. The MP2 reference converged for all twelve adsorbates.}
\end{deluxetable*}
\begin{deluxetable*}{llccc}
\tabletypesize{\scriptsize}
\tablecaption{Outlier basin analysis for adsorbates with pairwise method--method RMSDs greater than 1~\AA.\label{tab:outlier_analysis}}
\tablewidth{0pt}
\tablehead{
\colhead{Adsorbate} &
\colhead{Method} &
\colhead{$d_\mathrm{centroid}$ (\AA)} &
\colhead{Basin group} &
\colhead{RMSD vs. MP2 (\AA)}
}
\startdata
H$_2$CO & HF-3c/MINIX              & 4.52 & 1 & 0.054 \\
        & PBEh-3c/def2-mSVP        & 3.53 & 2 & 1.397 \\
        & r$^2$SCAN-3c/def2-mTZVPP & 3.47 & 2 & 1.403 \\
        & MPWB1K-D3BJ/def2-SVPD    & 4.37 & 1 & 0.091 \\
        & $\omega$B97X-D/def2-TZVP & 4.54 & 1 & 0.101 \\
        & MP2/def2-TZVP            & 4.45 & 1 & 0.000 \\
HCOOH   & HF-3c/MINIX              & 5.04 & 1 & 1.176 \\
        & PBEh-3c/def2-mSVP        & 5.32 & 1 & 1.258 \\
        & r$^2$SCAN-3c/def2-mTZVPP & 3.83 & 2 & 0.040 \\
        & MPWB1K-D3BJ/def2-SVPD    & 5.06 & 1 & 1.174 \\
        & $\omega$B97X-D/def2-TZVP & 3.90 & 2 & 0.030 \\
        & MP2/def2-TZVP            & 3.85 & 2 & 0.000 \\
\enddata
\tablecomments{Basin groups are assigned by single-link agglomeration with a 0.2\,\AA\ pairwise-RMSD threshold; $d_{\mathrm{centroid}}$ is the adsorbate--CO$_2$ centroid distance; RMSDs are reported relative to the MP2/def2-TZVP reference geometry. All twelve entries derive from converged geometry optimisations. For H$_2$CO, the production MPWB1K-D3BJ geometry agrees with the MP2 reference in Basin 1, while PBEh-3c and r$^2$SCAN-3c locate a closer-bound Basin 2. For HCOOH, the production MPWB1K-D3BJ geometry occupies Basin 1 together with HF-3c and PBEh-3c, while r$^2$SCAN-3c and $\omega$B97X-D reproduce the MP2 minimum in Basin 2; this is the only case in this dataset where the production and reference methods disagree on which of two bound configurations is the binding site.}
\end{deluxetable*}
\clearpage
\onecolumngrid
\subsection{Binding-energy benchmark}
\label{sec:be_benchmark}
\label{sec:be_benchmark_method}

To characterise the BE accuracy of density-functional approximations against a benchmark-quality wavefunction reference, we constructed a 36-system test set comprising three CP-corrected adsorbate--CO$_2$ cluster complexes per adsorbate on the CO$_2$ trimer, 19 on the cyclic CD${_3}$-01 isomer and 17 on the barrel CD${_3}$-02 isomer. The three geometries per adsorbate were selected from the production binding-site pool as the lowest-, median-, and highest-BE sites at the MPWB1K-D3BJ/def2-SVPD production level, so that the test set spans the within-cluster site-to-site BE spread rather than sampling only the deepest minimum. All complexes were evaluated as CP-corrected single points on the MPWB1K-D3BJ/def2-SVPD refinement geometry, isolating the functional dependence of the BE from any geometry dependence.

At each of the 36 systems, BEs were computed using 114 density-functional approximations with the def2-TZVP basis, covering every major class of functional, from simple to more sophisticated: range-separated hybrids ($\omega$B97X-V, $\omega$B97M-V, $\omega$B97X-D3BJ, $\omega$B97M-D3BJ, $\omega$B97X-D, CAM-B3LYP-D3BJ, $\omega$PBE-D3BJ, and variants); conventional global hybrids (PBE0-D3BJ, B3LYP-D3BJ, B1LYP-D3BJ, TPSSH-D3BJ, PW6B95-D3, MPW1B95-D3, and variants); Minnesota functionals (M05-2X, M06-2X, M06-HF, M08-SO, MN15, and variants); meta-GGAs and GGAs (PBE-D3BJ, BLYP-D3BJ, B97M-V, B97M-D3BJ, SCAN, SCAN-D3BJ, r$^2$SCAN, TPSS, M06-L, and variants); and double hybrids (DSD-BLYP-D3BJ and DSD-PBEP86-D3BJ) \citep{Mardirossian_2017}. The full ranking is given in Table~\ref{tab:be_bench_full}, in which the three production functionals retained for BE reporting ($\omega$B97X-V, $\omega$PBE-D3BJ, and M06-HF) are highlighted in bold.

Of the 4104 attempted single-point evaluations ($114 \times 36$), 3369 ($82.1\%$) yielded a converged SCF and contribute to the per-functional MAE statistics. The remaining $17.9\%$ failed to converge under the default Psi4 SCF settings, concentrated on the most strongly polar adsorbates: the strong H-bond donors CH$_3$OH, HCOOH and HNC, together with the strongly polar carbonyl H$_2$CO, account for the bulk of the failures, consistent with the larger electronic polarisation and steeper PES topology of these hydrogen-bonded complexes relative to the weak and dispersion-bound class. Per-functional MAEs were computed only over the systems for which the functional converged, with the converged count $N$ reported alongside each MAE so that small-$N$ rankings can be interpreted with appropriate caution.

For the wavefunction reference, we evaluated MP2 with the def2-TZVP, aug-cc-pVTZ, and aug-cc-pVQZ basis sets, and CCSD(T) with the aug-cc-pVTZ basis. A reference-quality CCSD(T)/CBS BE was then constructed using the focal-point approximation, combining the MP2 BE at the complete-basis-set (CBS) limit with the CCSD(T) correction evaluated at the aug-cc-pVTZ level:
\begin{equation}
\begin{gathered}
E_{\mathrm{b}}^{\mathrm{CCSD(T)/CBS}}
\approx
E_{\mathrm{b}}^{\mathrm{MP2/CBS}} 
+
\left[
E_{\mathrm{b}}^{\mathrm{CCSD(T)}}
-
E_{\mathrm{b}}^{\mathrm{MP2}}
\right]_{\mathrm{aug\text{-}cc\text{-}pVTZ}}
\end{gathered}
\label{eq:ccsdt_cbs_focal_point}
\end{equation}
where $E_{\mathrm{b}}$ is the BE, $E_{\mathrm{b}}^{\mathrm{MP2/CBS}}$ is the MP2 value at the CBS limit, and the term in parentheses is a correction from the more accurate CCSD(T) method, evaluated with the aug-cc-pVTZ basis. This construction combines two calculations that are each affordable on their own but would be far too expensive to perform together at full accuracy. MP2 is cheap enough to extrapolate to the basis-set limit, as described above. The CCSD(T) correction adds the electron-correlation effects that MP2 misses and can be evaluated with a smaller basis because this correction converges much faster with basis size than the underlying energy does. Adding the two gives our CCSD(T)/CBS reference at a fraction of the cost of a direct calculation. Even so, the CCSD(T) step dominates the expense: the complete reference set required 4569 core-hours in total, or 117 per binding site (range 37--260), of which 97\% was consumed by the CCSD(T)/aug-cc-pVTZ calculations.

Per-functional statistics relative to this reference are given in Table~\ref{tab:be_bench_full}. Of the 114 density functionals tested, $\omega$B97X-V/def2-TZVP delivers the smallest MAE ($= 53$\,K over $N = 33$ converged systems out of 36), a near-zero mean error (ME; $= +6$\,K), and an RMSE of 72\,K, with a maximum absolute deviation of 214\,K for any single converged system. This 53\,K MAE sets the smallest uncertainty ($\sigma$) adopted throughout the per-cluster BE results (Table~\ref{tab:final_BE}, Figures~\ref{fig:be_boxplots} and~\ref{fig:be_convergence}). Four of the top five functionals belong to the same family (the $\omega$B97 range-separated hybrids: $\omega$B97X-V, $\omega$B97M-V, $\omega$B97X-D3BJ, and $\omega$B97M-D3BJ), agreeing to within 13\,K of one another. These functionals share two features well suited to binding to CO$_2$: range-separated exchange, which gives the correct asymptotic behaviour of the exchange potential, and an explicit account of dispersion, treated nonlocally through VV10 in $\omega$B97X-V and $\omega$B97M-V and by the atom-pairwise D3BJ scheme in the other two. The two dispersion models agree to within 13\,K here, so the result does not hinge on which is used. A similar ordering has been reported for water--CO$_2$ clusters, where $\omega$B97X reproduces CCSD(T)/CBS binding energies closely while $\omega$B97X-D and B3LYP underbind substantially \citep{Thanthiriwatte_2012}.

The benchmark is carried out on the trimer, using both the cyclic CD$_3$-01 and barrel CD$_3$-02 isomers. It therefore does not capture an effect that becomes important on larger clusters: when several CO$_2$ molecules surround the adsorbate, their combined dispersion interaction can be stronger than the sum of the individual adsorbate--CO$_2$ interactions. This additional many-body contribution is described differently by different functionals, so the reported BEs retain some dependence on the functional used.

To test how sensitive the results are to the choice of functional, we recomputed every BE with two additional functionals. Both predict values that are lower than those obtained with $\omega$B97X-V by about 250\,K on average. This reflects the tendency of $\omega$PBE-D3BJ and M06-HF to underbind dispersion-dominated complexes, rather than an uncertainty in the reported values. Against the CCSD(T)/CBS reference, $\omega$B97X-V has a mean signed error of $+6$\,K and an RMSE of 72\,K, compared with $+209$\,K and $+220$\,K for $\omega$PBE-D3BJ and M06-HF, respectively. We therefore base the reported BEs on $\omega$B97X-V, which has an MAE of 53\,K per site. The tetramer-to-pentamer trend agrees across all three methods for ten of the twelve adsorbates. The exceptions are N$_2$ ($+77$\,K at $\omega$B97X-V against $-13$ and $-30$\,K at $\omega$PBE-D3BJ and M06-HF) and HCl ($+57$ and $+55$\,K at $\omega$B97X-V and $\omega$PBE-D3BJ against $-97$\,K at M06-HF). All of these changes are below 100\,K and within the benchmark MAEs of $\omega$PBE-D3BJ and M06-HF (211 and 255\,K). The additional functionals are retained as a check that the main trend is robust.

The CBS extrapolation is needed for a reliable reference, as the same calculations without it differ by $\sim 100$\,K. We also tested MP2 as a cheaper alternative. At the same basis-set limit, MP2 reproduces the CCSD(T)/CBS BEs to within 81\,K, similar to the best-performing functionals. This agreement is largely due to error cancellation: MP2 overestimates dispersion and underestimates induction, the response of a molecule's electrons to the electric field of a neighbour. These errors offset each other for the H-bonded systems studied here, but not in a systematic way, so MP2 should be used with caution for this class of system.
\startlongtable
\begin{deluxetable*}{rlcrrrr}
\tabletypesize{\scriptsize}
\tablewidth{0pt}

\tablecaption{Full 114-functional binding energy benchmark ranking.
\label{tab:be_bench_full}}

\tablehead{
\colhead{Rank} &
\colhead{Functional / def2-TZVP} &
\colhead{$N$} &
\colhead{ME} &
\colhead{MAE} &
\colhead{RMSE} &
\colhead{max$|\Delta|$} \\
\colhead{} &
\colhead{} &
\colhead{} &
\colhead{(K)} &
\colhead{(K)} &
\colhead{(K)} &
\colhead{(K)}
}

\startdata
\textbf{1} & \textbf{$\omega$B97X-V} & \textbf{33} & \textbf{+6} & \textbf{53} & \textbf{72} & \textbf{214} \\
2 & PWB6K-D3BJ & 25 & $-22$ & 56 & 90 & 259 \\
3 & $\omega$B97M-D3BJ & 33 & $+45$ & 63 & 85 & 226 \\
4 & $\omega$B97M-V & 34 & $+2$ & 64 & 91 & 308 \\
5 & $\omega$B97X-D3BJ & 32 & $-1$ & 66 & 88 & 233 \\
6 & B97M-D3BJ & 27 & $-7$ & 66 & 83 & 197 \\
7 & B3LYP-D3BJ & 31 & $+6$ & 69 & 89 & 198 \\
8 & N12-SX-D3BJ & 29 & $-20$ & 76 & 94 & 195 \\
9 & PBE0-D3BJ & 32 & $+64$ & 80 & 120 & 333 \\
10 & B1LYP-D3BJ & 34 & $-38$ & 82 & 119 & 371 \\
11 & B97M-V & 24 & $+26$ & 84 & 104 & 227 \\
12 & PW6B95-D3 & 24 & $+36$ & 85 & 108 & 252 \\
13 & MPW1B95-D3 & 25 & $+57$ & 95 & 122 & 331 \\
14 & CAM-B3LYP-D3BJ & 32 & $-94$ & 102 & 144 & 405 \\
15 & $\omega$B97X & 32 & $-37$ & 106 & 156 & 518 \\
16 & $\omega$PBE-D3 & 31 & $+110$ & 112 & 145 & 351 \\
17 & MPWB1K-D3BJ & 25 & $+106$ & 115 & 150 & 412 \\
18 & HSE06-D3BJ & 33 & $+102$ & 116 & 172 & 588 \\
19 & PBE-D3BJ & 25 & $+104$ & 124 & 228 & 836 \\
20 & PW6B95-D3BJ & 26 & $+128$ & 133 & 185 & 475 \\
21 & DSD-BLYP-D3BJ & 33 & $+133$ & 133 & 169 & 486 \\
22 & X3LYP-D3BJ & 30 & $-135$ & 136 & 175 & 405 \\
23 & MN15 & 32 & $-18$ & 139 & 182 & 406 \\
24 & M05-2X & 29 & $+33$ & 141 & 172 & 428 \\
25 & $\omega$PBE-D3MBJ & 33 & $+140$ & 141 & 177 & 419 \\
26 & M08-SO & 35 & $+24$ & 157 & 198 & 502 \\
27 & DSD-PBEP86-D3BJ & 33 & $+159$ & 159 & 208 & 643 \\
28 & SCAN & 28 & $+63$ & 160 & 197 & 475 \\
29 & SCAN-D3BJ & 28 & $-125$ & 161 & 235 & 668 \\
30 & M06-2X & 32 & $+10$ & 161 & 216 & 599 \\
31 & MN12-SX-D3BJ & 29 & $+137$ & 162 & 205 & 385 \\
32 & DSD-PBEB95-D3BJ & 28 & $+170$ & 170 & 212 & 587 \\
33 & M11-D3BJ & 35 & $+1$ & 171 & 223 & 535 \\
34 & BLYP-D3BJ & 25 & $+164$ & 171 & 246 & 752 \\
35 & M08-HX & 31 & $+50$ & 177 & 216 & 513 \\
36 & B2PLYP-D3BJ & 33 & $+185$ & 185 & 232 & 651 \\
37 & CAM-B3LYP-D3 & 31 & $-190$ & 194 & 274 & 787 \\
38 & TPSSH-D3 & 28 & $+180$ & 194 & 281 & 848 \\
39 & SCAN0 & 32 & $+124$ & 199 & 218 & 357 \\
40 & MPW1B95-D3BJ & 25 & $+196$ & 200 & 235 & 500 \\
41 & PWB6K & 24 & $+188$ & 208 & 243 & 462 \\
42 & SOGGA11-X-D3BJ & 33 & $+145$ & 208 & 259 & 550 \\
\textbf{43} & \textbf{$\omega$PBE-D3BJ} & \textbf{32} & \textbf{+209} & \textbf{211} & \textbf{267} & \textbf{614} \\
44 & WB97X-D3 & 32 & $+211$ & 212 & 249 & 591 \\
45 & REVTPSSH-D3BJ & 31 & $+212$ & 217 & 299 & 797 \\
46 & M11 & 35 & $+123$ & 217 & 251 & 465 \\
47 & RPBE-D3BJ & 20 & $+179$ & 221 & 380 & 1117 \\
48 & $\omega$B97 & 32 & $-155$ & 228 & 313 & 899 \\
49 & MN15-L & 27 & $-113$ & 234 & 308 & 803 \\
50 & TPSS-D3BJ & 26 & $+225$ & 234 & 360 & 1188 \\
51 & B97-2-D3BJ & 33 & $+207$ & 237 & 320 & 759 \\
52 & r$^2$SCAN & 29 & $+216$ & 244 & 278 & 535 \\
53 & REVPBE0-D3BJ & 31 & $+241$ & 244 & 334 & 779 \\
54 & PTPSS-D3BJ & 34 & $+251$ & 251 & 297 & 705 \\
\textbf{55} & \textbf{M06-HF} & \textbf{34} & \textbf{+220} & \textbf{255} & \textbf{310} & \textbf{694} \\
56 & TPSSH-D3BJ & 29 & $+247$ & 255 & 371 & 1022 \\
57 & PWPB95-D3BJ & 28 & $+267$ & 267 & 333 & 798 \\
58 & BMK-D3BJ & 32 & $+169$ & 273 & 332 & 755 \\
59 & M06-L & 27 & $+283$ & 293 & 337 & 669 \\
60 & B97-D3BJ & 26 & $+286$ & 295 & 467 & 1396 \\
61 & revPBE-D3BJ & 22 & $+305$ & 319 & 500 & 1497 \\
62 & M05 & 31 & $+334$ & 334 & 418 & 1225 \\
63 & MGGA\_MS2H & 32 & $+337$ & 337 & 414 & 1012 \\
64 & MN12-SX & 29 & $+345$ & 345 & 391 & 653 \\
65 & BP86-D3BJ & 25 & $+356$ & 359 & 452 & 1078 \\
66 & SOGGA & 19 & $+366$ & 366 & 480 & 1254 \\
67 & M06 & 30 & $+373$ & 373 & 408 & 835 \\
68 & $\omega$B97X-D & 34 & $+386$ & 387 & 452 & 996 \\
69 & CAM-B3LYP & 32 & $+432$ & 432 & 532 & 1047 \\
70 & B97-D & 26 & $+477$ & 482 & 649 & 1867 \\
71 & MPWB1K & 25 & $+484$ & 484 & 547 & 1093 \\
72 & PW91 & 24 & $+488$ & 488 & 676 & 1970 \\
73 & MPW1LYP & 34 & $+488$ & 488 & 617 & 1494 \\
74 & MN12-L & 28 & $+512$ & 512 & 553 & 824 \\
75 & PW6B95 & 25 & $+517$ & 517 & 597 & 1261 \\
76 & BHandHLYP & 33 & $+527$ & 527 & 646 & 1244 \\
77 & B97-1 & 33 & $+529$ & 529 & 649 & 1681 \\
78 & N12-SX & 29 & $+549$ & 549 & 637 & 1194 \\
79 & PWPB95 & 28 & $+550$ & 550 & 644 & 1404 \\
80 & MPW1B95 & 25 & $+564$ & 564 & 638 & 1048 \\
81 & HSE03 & 31 & $+595$ & 595 & 715 & 1647 \\
82 & PBE0 & 32 & $+603$ & 603 & 704 & 1344 \\
83 & HSE06 & 33 & $+607$ & 607 & 717 & 1713 \\
84 & PTPSS & 34 & $+619$ & 618 & 711 & 1542 \\
85 & PBE & 25 & $+651$ & 651 & 828 & 2228 \\
86 & B2PLYP & 33 & $+661$ & 661 & 769 & 1726 \\
87 & M11-L & 18 & $+662$ & 662 & 721 & 1115 \\
88 & SOGGA11 & 30 & $+280$ & 693 & 965 & 2806 \\
89 & X3LYP & 30 & $+710$ & 710 & 848 & 1854 \\
90 & MPW1K & 31 & $+758$ & 758 & 879 & 1670 \\
91 & LRC-WPBEH & 35 & $+765$ & 765 & 906 & 2043 \\
92 & SOGGA11-X & 33 & $+830$ & 830 & 919 & 1602 \\
93 & BMK & 32 & $+840$ & 840 & 923 & 1463 \\
94 & revTPSSH & 31 & $+870$ & 870 & 1031 & 2345 \\
95 & LRC-WPBE & 35 & $+898$ & 898 & 1056 & 2354 \\
96 & BHANDH & 34 & $-890$ & 901 & 1151 & 2546 \\
97 & TPSS & 27 & $+920$ & 920 & 1158 & 2994 \\
98 & B1LYP & 33 & $+946$ & 945 & 1106 & 2354 \\
99 & TPSSH & 29 & $+952$ & 952 & 1173 & 2757 \\
100 & MPW1PBE & 33 & $+955$ & 955 & 1105 & 2092 \\
101 & B3LYP & 31 & $+955$ & 955 & 1133 & 2412 \\
102 & B97-3 & 32 & $+979$ & 979 & 1119 & 2393 \\
103 & MPW1PW91 & 33 & $+997$ & 997 & 1155 & 2534 \\
104 & B1B95 & 19 & $+1009$ & 1009 & 1175 & 2408 \\
105 & B3P86 & 31 & $+1043$ & 1043 & 1188 & 2435 \\
106 & BLYP & 24 & $+1129$ & 1129 & 1381 & 3430 \\
107 & N12 & 30 & $+1193$ & 1193 & 1397 & 2861 \\
108 & RPBE & 20 & $+1255$ & 1255 & 1653 & 4144 \\
109 & B97-2 & 33 & $+1259$ & 1259 & 1481 & 3230 \\
110 & REVPBE0 & 31 & $+1263$ & 1263 & 1457 & 2708 \\
111 & BP86 & 25 & $+1291$ & 1291 & 1531 & 3344 \\
112 & B1PW91 & 31 & $+1358$ & 1358 & 1519 & 2577 \\
113 & B3PW91 & 32 & $+1368$ & 1368 & 1585 & 3302 \\
114 & revPBE & 22 & $+1383$ & 1383 & 1790 & 4443 \\
\enddata
{\tablecomments{Functionals are benchmarked against the CCSD(T)/CBS reference on a 36-system test set comprising 12 adsorbates and three trimer binding sites per adsorbate (19 on CD${_3}$-01 and 17 on CD${_3}$-02). Each adsorbate--site pair is treated as an independent test-set entry. Statistics are computed over the $N$ converged entries, out of 36, for which each functional yielded a converged SCF. The ME is positive when the functional underbinds relative to CCSD(T)/CBS. The three production functionals retained in the BE reporting pipeline ($\omega$B97X-V/def2-TZVP, $\omega$PBE-D3BJ/def2-TZVP, and M06-HF/def2-TZVP) were selected to span complementary physical approximations of non-covalent interactions rather than by pooled MAE alone and are highlighted in bold.}}
\end{deluxetable*}
\begin{deluxetable*}{lcccccccccccc}
\tablecaption{Mean binding energies (K) of the twelve adsorbates at the three production functionals on CD${_2}$--CD${_5}$. Trimer columns pool the cyclic (CD${_3}$-01) and barrel (CD${_3}$-02) isomers. \label{tab:be_three_functional}}
\tablewidth{0pt}
\tablehead{
\colhead{Species} &
\colhead{$\omega$B97X-V} &
\colhead{$\omega$PBE-D3BJ} &
\colhead{M06-HF} &
\colhead{$\omega$B97X-V} &
\colhead{$\omega$PBE-D3BJ} &
\colhead{M06-HF} &
\colhead{$\omega$B97X-V} &
\colhead{$\omega$PBE-D3BJ} &
\colhead{M06-HF} &
\colhead{$\omega$B97X-V} &
\colhead{$\omega$PBE-D3BJ} &
\colhead{M06-HF} \\
&
\colhead{CD2} &
\colhead{CD2} &
\colhead{CD2} &
\colhead{CD3} &
\colhead{CD3} &
\colhead{CD3} &
\colhead{CD4} &
\colhead{CD4} &
\colhead{CD4} &
\colhead{CD5} &
\colhead{CD5} &
\colhead{CD5}
}
\startdata
$\mathrm{N_2}$     & 275 & 183 & 138 & 365 & 262 & 54 & 486 & 401 & 142 & 563 & 387 & 112 \\
$\mathrm{CH_4}$    & 302 & 243 & 104 & 198 & 154 & 23 & 611 & 471 & 298 & 331 & 203 & -28 \\
$\mathrm{HCl}$     & 557 & 472 & 295 & 799 & 673 & 415 & 1554 & 1364 & 1102 & 1612 & 1419 & 1005 \\
$\mathrm{C_2H_2}$  & 815 & 736 & 685 & 939 & 845 & 790 & 1865 & 1719 & 1705 & 1763 & 1589 & 1550 \\
$\mathrm{H_2S}$    & 754 & 709 & 602 & 1046 & 979 & 867 & 1008 & 948 & 749 & 1385 & 1307 & 1269 \\
$\mathrm{HF}$      & 783 & 630 & 478 & 2030 & 1576 & 1777 & 2368 & 1752 & 2234 & 2590 & 2023 & 2332 \\
$\mathrm{HCN}$     & 1340 & 1147 & 1215 & 1505 & 1298 & 1224 & 1690 & 1530 & 1276 & 2216 & 1933 & 1818 \\
$\mathrm{H_2CO}$   & 638 & 500 & 567 & 827 & 655 & 714 & 1512 & 1212 & 1374 & 1831 & 1491 & 1492 \\
$\mathrm{HCOOH}$   & 1345 & 1109 & 1153 & 2401 & 2003 & 2189 & 3172 & 2643 & 2890 & 4364 & 3689 & 3941 \\
$\mathrm{HNC}$     & 1213 & 1094 & 774 & 1553 & 1416 & 1153 & 1868 & 1710 & 1457 & 2065 & 1924 & 1667 \\
$\mathrm{CH_3OH}$  & 1303 & 1095 & 1046 & 1934 & 1605 & 1894 & 3327 & 2814 & 3413 & 3410 & 2861 & 3523 \\
$\mathrm{H_2O}$    & 1468 & 1177 & 1508 & 1572 & 1260 & 1516 & 1613 & 1294 & 1710 & 2689 & 2218 & 2583 \\
\enddata
\tablecomments{The pentamer values for N$_2$, CH$_4$, HCl, C$_2$H$_2$, H$_2$CO and H$_2$O were computed on BLYP-D3BJ/def2-SVPD geometries, following Section~\ref{subsec:rescue_protocol}.}
\end{deluxetable*}

\end{document}